\documentclass[11pt]{article}
\usepackage[centertags]{amsmath}
\usepackage[square, comma, sort&compress,numbers]{natbib}
\usepackage{array,multirow}
\numberwithin{equation}{section}
\usepackage{amssymb}
\usepackage{graphicx}
\usepackage{color}
\usepackage{epsfig}
\usepackage{bbold}

\usepackage{wrapfig}
\usepackage{float}
\usepackage{soul}

\usepackage{tkz-euclide}

\usepackage{tikz,pgf}
\usetikzlibrary{shapes}
\usetikzlibrary{calc}
\usetikzlibrary{decorations.pathmorphing}
\usetikzlibrary{decorations.pathreplacing,shapes.misc}
\usetikzlibrary{positioning}
\usetikzlibrary{arrows}
\usetikzlibrary{decorations.markings}
\usetikzlibrary{shadings}
\usetkzobj{all}

\usetikzlibrary{intersections}

\def\be{\begin{equation}}
\def\ee{\end{equation}}
\def\ba{\begin{array}}
\def\ea{\end{array}}

\def\dps{\displaystyle}

\def\1{\tilde{1}}
\def\2{\tilde{2}}
\def\3{\tilde{3}}

\newdimen\tableauside\tableauside=1.0ex
\newdimen\tableaurule\tableaurule=0.4pt
\newdimen\tableaustep
\def\phantomhrule#1{\hbox{\vbox to0pt{\hrule height\tableaurule
width#1\vss}}}
\def\phantomvrule#1{\vbox{\hbox to0pt{\vrule width\tableaurule
height#1\hss}}}
\def\sqr{\vbox{%
  \phantomhrule\tableaustep

\hbox{\phantomvrule\tableaustep\kern\tableaustep\phantomvrule\tableaustep}%
  \hbox{\vbox{\phantomhrule\tableauside}\kern-\tableaurule}}}
\def\squares#1{\hbox{\count0=#1\noindent\loop\sqr
  \advance\count0 by-1 \ifnum\count0>0\repeat}}
\def\tableau#1{\vcenter{\offinterlineskip
  \tableaustep=\tableauside\advance\tableaustep by-\tableaurule
  \kern\normallineskip\hbox
    {\kern\normallineskip\vbox
      {\gettableau#1 0 }%
     \kern\normallineskip\kern\tableaurule}%
  \kern\normallineskip\kern\tableaurule}}
\def\gettableau#1 {\ifnum#1=0\let\next=\null\else
  \squares{#1}\let\next=\gettableau\fi\next}

\newcommand{\bref}[1]{\textbf{\ref{#1}}}

\def\cF{\mathcal{F}}

\def\cO{\mathcal{O}}

\numberwithin{equation}{section} \makeatletter
\@addtoreset{equation}{section}

\def\ads{AdS_{3}}
\def\cft{CFT_{2}}
\def\dual{\ads/\cft}
\def\be{\begin{equation}}
\def\ee{\end{equation}}
\def\ba{\begin{array}}
\def\ea{\end{array}}

\def\dps{\displaystyle}

\def\ba{\begin{array}}
\def\ea{\end{array}}

\def\dps{\displaystyle}

\def\tepsilon{\tilde\epsilon}
\def\bnd{\partial \mathbb{D}}
\def\disk{\mathbb{D}}
\def\wdisk{\disk_\alpha}
\def\wads{\ads^{(\alpha)}}

\usepackage{jheppub}
\usepackage{amsmath,amsfonts,amssymb,amsthm}

\makeatletter
\def\@fpheader{\vspace{-.1cm}}
\makeatother

\title{Perturbative classical conformal blocks as Steiner trees \\  on the hyperbolic disk}

\author[a,b]{Konstantin\ Alkalaev}

\author[a]{Mikhail\ Pavlov}

\affiliation[a]{I.E. Tamm Department of Theoretical Physics, \\P.N. Lebedev Physical
Institute,\\ Leninsky ave. 53, 119991 Moscow, Russia}
\affiliation[b]{Department of General and Applied Physics, \\
Moscow Institute of Physics and Technology, \\
Institutskiy per. 7, Dolgoprudnyi, \\141700 Moscow region, Russia}
\emailAdd{alkalaev@lpi.ru}
\emailAdd{pavlov@lpi.ru}

\abstract{We consider the Steiner tree problem in hyperbolic geometry in the context of the  AdS/CFT duality between large-$c$ conformal blocks on the boundary and particle motions in the bulk. The Steiner trees  are weighted graphs on the Poincare disk with a number of endpoints and  trivalent vertices connected to each other by edges in such a way that an overall length is minimum. We specify a particular class of  Steiner trees that we call  holographic. Their characteristic property is that a holographic Steiner tree with $N$ endpoints can be inscribed into an $N$-gon with $N-1$ ideal vertices. The holographic Steiner trees are dual to  large-$c$ conformal blocks. Particular examples of $N=2,3,4$ Steiner trees as well as their dual conformal blocks are explicitly calculated. We discuss  geometric properties of the holographic Steiner trees and their realization in CFT terms. It is shown that connectivity and cuts of the Steiner trees encode the factorization properties of large-$c$ conformal blocks.
}

\preprint{FIAN-TD-2018-20}
\arxivnumber{}

\begin{document}

\maketitle
\flushbottom

\section{Introduction}

The recent study of the  $\dual$ correspondence \cite{Brown:1986nw,Aharony:1999ti} in the large-$c$ regime has brought many novel insights and fruitful observations, of which the most important is that conformal blocks of boundary CFT have gained an independent holographic interpretation. In particular, large-$c$ conformal blocks associated to light and heavy primary operators can be independently described as lengths of particular geodesic networks in the bulk space \cite{Hartman:2013mia,Fitzpatrick:2014vua,Caputa:2014eta,Hijano:2015rla,Fitzpatrick:2015zha,Alkalaev:2015wia,Hijano:2015qja,Alkalaev:2015lca,Fitzpatrick:2015dlt,Banerjee:2016qca,Chen:2016dfb,Alkalaev:2016rjl}.\footnote{For further development see e.g. \cite{Kraus:2016nwo,Hulik:2016ifr,Fitzpatrick:2016mtp,Belavin:2017atm,Maxfield:2017rkn,Kusuki:2018wcv,Brehm:2018ipf,Kusuki:2018nms}. The study of semiclassical conformal blocks in the holographic context can be extended in many directions including any dimensions, $1/c$ corrections, $W_N$ symmetry,  other topologies, Gallilean symmetry, BMS$_3$, and supersymmetric extensions   \cite{Fitzpatrick:2014oza,deBoer:2014sna,Bobev:2015jxa,Hijano:2015zsa,Beccaria:2015shq,Alkalaev:2016ptm,Alkalaev:2016fok,Bagchi:2017cpu,Kraus:2017ezw,Alkalaev:2017bzx,Menotti:2018jsy,Alkalaev:2018qaz,Belavin:2018hfm,Hikida:2018eih,Bombini:2018jrg,Lodato:2018gyp,Hulik:2018dpl}.} These developments highlighted the fundamental role of conformal blocks in holographic duality that was previously poorly understood.

In general terms, the conformal blocks of two-dimensional CFT with Virasoro symmetry are (anti)ho\-lo\-mor\-phic functions on Riemannian surfaces additionally parameterized by external and intermediate conformal dimensions $\Delta$ and the central charge $c$ \cite{Belavin:1984vu}. The block functions are unknown in a closed form, however, using various large-$c$ approximations in the parameter space ($\Delta, c$) their form can be essentially simplified. The most striking example here are the global or light blocks (depending on the surface's genus) that are obtained as limiting case $c\to \infty$ of the original conformal blocks with fixed dimensions $\Delta = \cO(c^0)$, see e.g. \cite{Belavin:1984vu,ZZbook,Perlmutter:2015iya,Alkalaev:2015fbw,Alkalaev:2016fok,Cho:2017oxl,Rosenhaus:2018zqn}. The other seminal example of approximate conformal block functions is the classical conformal block \cite{Zamolodchikov1986} that arises in the large-$c$ regime with  linearly growing dimensions  $\Delta = \cO(c^1)$, see e.g. \cite{Piatek:2013ifa,Hijano:2015rla,Alkalaev:2015wia}. There are also plenty of heavy-light conformal blocks with conformal dimensions behaving like $\cO(c^0)$ (light operator) and/or $\cO(c^1)$ (heavy operator), see e.g. \cite{Fitzpatrick:2014vua,Hijano:2015rla,Fitzpatrick:2015zha,Alkalaev:2015wia,Belavin:2017atm}. It turns out that these approximate conformal blocks are not independent and can be related to each other by different maps \cite{Fitzpatrick:2015zha,Alkalaev:2015fbw,Alkalaev:2016fok}.

In this context, the most studied case of the semiclassical $\dual$ correspondence is when there are two heavy primary operators in the boundary CFT that produce an angle deficit or BTZ block hole in the bulk space \cite{Fitzpatrick:2014vua}. Here, the classical conformal blocks can be calculated using the so-called heavy-light perturbation theory. The dual description reveals a number of interacting point massive particles (with masses $\sim \Delta/c$) propagating on the background and the total length of their worldlines is exactly the perturbative classical conformal block.

Explicit calculation of dual functions which are perturbative classical blocks and lengths of the worldline networks is both conceptually and technically complicated problem. On the boundary, one can use the monodromy method of calculating $n$-point conformal blocks that relies on solving higher order polynomial equation systems  \cite{Hartman:2013mia,Fitzpatrick:2014vua,Hijano:2015rla,Banerjee:2016qca,Alkalaev:2016rjl}. Up to now, exact expressions were known only for $n=4$ blocks because the $n$-point monodromy equations reduce to $(2n-6)$-th order polynomial equation.
In the bulk, in order to calculate the total length of the network one can use the worldline formalism for classical mechanics of massive particles \cite{Hijano:2015rla,Alkalaev:2015wia}. This approach is also hard to implement in the higher point case, and, therefore, the problem of calculating network lengths remains open. Nonetheless, one can prove the existence theorem claiming that dual  functions  exist and equal to each other in the $n$-point case for any $n$ \cite{Alkalaev:2016rjl}.

In this paper we develop new  techniques employing a purely geometric view of the bulk dynamics. To this end, it is convenient to represent the original bulk space in $z,\bar z$ and $t$ coordinates and then fix a constant time $t=0$ slice. In what follows we focus on the bulk space with a conical singularity. In this case, having an angle deficit in $\ads$ spacetime we obtain  the Poincare disk $\disk$ with the same defect. Let us recall now that geodesics on the hyperbolic disk are represented by  half-circles perpendicular to the (conformal) boundary circle. Then, discarding any mechanical interpretation of the bulk dynamics  we can reformulate the whole problem  in terms of the Poincare disk model geometry.

We observe that interacting particle worldlines  we are interested in are exactly the so-called {\it Steiner trees} in  metric spaces  (for review see e.g. \cite{Ivanov_Tuzhilin, Courant}). The Steiner tree problem in graph theory is the optimization problem of finding  shortest path along a particular graph with a given number of endpoints, edges, and vertices. In the context of the $\dual$ correspondence  we deal with particular Steiner trees  in two-dimensional hyperbolic geometry that we call {\it holographic} Steiner trees. To establish correspondence with conformal blocks we calculate the lengths of holographic Steiner trees on the Poincare disk. Note that in this setting the problem  is purely mathematical without any direct reference to its original physical motivation.

Identifying perturbative classical blocks with holographic Steiner trees we reveal one more interesting aspect. Recall that in the monodromy method $n$-point  block functions are defined via accessory parameters subject to the system of $n$ quadratic equations. Presently, it is not clear  whether the equations are solvable or not in the sense of Galois theory. Then, the geometrical method based on Steiner trees on the Poincare disk, and, more broadly, the semiclassical $AdS_3/CFT_2$ correspondence,  can be viewed as the associated compass-and-straightedge construction to find roots of  algebraic equations.

The paper is organized as follows. In the first part of the paper we introduce Steiner trees and describe  their geometric properties in the bulk, while in the second part we turn to the boundary CFT analysis and discuss the monodromy method for calculating classical conformal blocks. In both parts we focus on  the structures that help to underline the duality between two descriptions.  In Section \bref{sec:geometry} we describe basic geometric facts about Steiner trees both on Euclidean and hyperbolic planes. In Section \bref{sec:examples}  several examples of holographic Steiner trees with at most two vertices are calculated. Section \bref{sec:perturbative} contains a general view of perturbative classical conformal blocks. We propose a slight technical  modification of the standard monodromy method where the coordinate of the first primary operator $z_1 \neq 0$. Here we also discuss the correspondence formula between blocks and lengths. In Section \bref{sec:lower} we explicitly calculate various lower point conformal blocks and compare the resulting expressions with the Steiner tree analysis of the first part.   Section \bref{sec:factor} is central in our  analysis of dual conformal blocks, it classifies  identity conformal blocks and proves their factorization properties.  Technical details are given in Appendices \bref{app:A}, \bref{app:N34}, \bref{app:factor}.

\section{Steiner trees on the hyperbolic disk}
\label{sec:geometry}

The Steiner tree problem in graph theory is the optimization problem of finding  shortest path along a particular graph with a given number of endpoints, edges, and vertices (for review and mathematically rigorous treatment see, e.g., \cite{Ivanov_Tuzhilin, Courant}). In what follows we  introduce Steiner trees on the (non-)Euclidean plane and discuss their properties relevant from the holographic duality perspective.

\subsection{Steiner trees and Fermat-Torricelli  points}

Let $N$ be a number of points on the (non)-Euclidean plane. The points are connected to each other by edges so that there are $N-2$ trivalent vertices. Suppose that each edge carries a weight $\epsilon \in \mathbb{R}$.  The resulting graph is required to have a minimum total length
\be
\label{minimum}
\hspace{-15mm}\text{shortest path}\;:\qquad L_N = \sum_{a\in \{\text{edges}\}} \epsilon_a L_a\;,
\ee
where $\epsilon_a$ are the edge weights and $L_a$ are lengths of edges calculated using  the corresponding metric. Positions of vertices are fixed by \eqref{minimum} in terms of  weights and endpoints. In this way we obtain a weighted Steiner tree.

\begin{figure}[H]
\centering
\includegraphics[width=90mm]{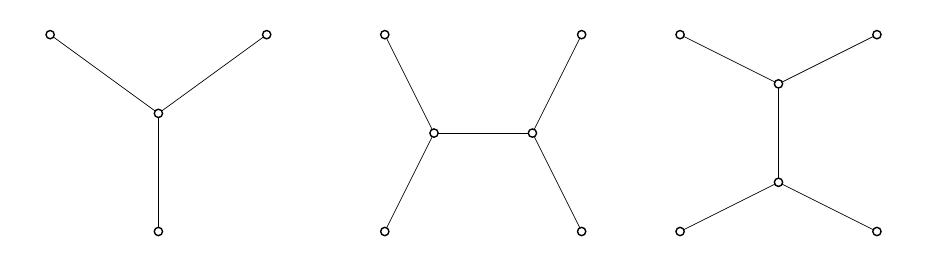}
\caption{Simplest Euclidean Steiner trees with equal weights: (a) $N=3$ tree with a single vertex known as the  Fermat-Torricelli point, it is inscribed into a triangle, (b) $N=4$ tree with two vertices inscribed into a square. (This case was discussed by Courant and Robbins \cite{Courant}. Curiously,  the two Steiner trees look like two OPE channels of the 4-point conformal block.)}
\label{euclid}
\end{figure}

The most striking property of Steiner trees originally shown on the euclidean plane is that vertices are the Fermat-Torricelli (FT) points where edges form three $120^\circ$ angles, see Fig. \bref{euclid}. Recall that the Fermat-Torricelli point of a triangle is defined as a point such that the total length from the three vertices of the triangle to the point is minimal.

This observation is naturally extended to arbitrary edge weights so that the angles are given by
\be
\label{cooos}
  \ba{c}
      \dps \cos \gamma_{ca} = \frac{\epsilon^2_a-\epsilon^2_c - \epsilon^2_b}{2 \epsilon_a \epsilon_c}\;,\qquad
      \dps \cos \gamma_{bc} = \frac{\epsilon^2_b-\epsilon^2_c - \epsilon^2_a  }{2 \epsilon_b\epsilon_c }\;,
      \qquad
       \dps \cos \gamma_{ab} = \frac{\epsilon^2_c - \epsilon^2_b - \epsilon^2_a}{2 \epsilon_a \epsilon_b}\;.
  \ea
\ee
Here labels $a,b,c$ enumerate three edges forming a vertex, $\epsilon_{a,b,c}$ are the edge weights, and $\gamma_{ab}, \gamma_{bc}, \gamma_{ca}$ are the angles between the respective edges (see Fig. \bref{vertex}). Obviously, $\gamma_{ab}+\gamma_{bc}+\gamma_{ca} = 2\pi$.  It follows that the edge weights necessarily satisfy the triangle inequalities

\be
\label{triangle}
\ba{c}
\epsilon_a + \epsilon_b \geq \epsilon_c\;,
\qquad
\epsilon_a + \epsilon_c \geq \epsilon_b\;,
\qquad
\epsilon_b + \epsilon_c \geq \epsilon_a\;.
\ea
\ee
Relations \eqref{cooos} and \eqref{triangle} follow from the minimum total length requirement \eqref{minimum}. The cosine formulas \eqref{cooos} and the triangle inequalities \eqref{triangle} were previously discussed in the mathematical literature \cite{10.2307/2695644,Link,Zachos}. In the context of the block/length correspondence these formulas were found  in \cite{Alkalaev:2016rjl}.

\subsection{Poincare disk model}
\label{sec:disk}

A time slice of the $AdS_3$ spacetime is given by the two-dimensional hyperbolic plane that can be described within the Poincare disk model. Considering the $AdS_3$ spacetime with an angle deficit parameterized by $\alpha$ that we denote as $\wads$ results in cutting a wedge out of the disk. Then, the metric reads
\be
\label{ss1_2}
ds^2=\frac{4dz d\bar{z}}{(1-z\bar{z})^2} \;,
\ee
where coordinates $z = r \exp[i\omega]$ with $r\in [0,1)$ and $\omega \in [0,2\pi\alpha)$ cover a disk $\mathbb{D}$ with an angle deficit parameterized by $\alpha \in (0,1]$.  The conformal boundary $\partial \mathbb{D}$ is described by (a part of) the boundary circle $z\bar z = 1$. The disk with the angle deficit parameter $\alpha$ and its  boundary circle  will be denoted as $\wdisk$ and $\partial \wdisk$.

Rescaling the angular coordinate on the disk as $\omega\to \alpha\, \omega$ we reproduce the  Poincare disk model of the hyperbolic geometry. We shall use this fact to simplify our analysis of geodesics by calculating their lengths on the Poincare disk first and then rescaling by $\alpha$ all angular coordinates. This is legitimate because Steiner  trees we consider depend on endpoints $w \in \partial \mathbb{D}$ only (see below). From now on we work with the Poincare disk model and restore the $\alpha$-dependence in final expressions.

It is useful to recall that the isometry of the Poincare disk $\mathbb{D}$ is given by the  M{\"o}bius transformations
\be
\label{mob}
z \to \frac{a z + b}{\bar b z +\bar{a}}\;,\quad a,b \in \mathbb{C}\;, \quad |a|^2-|b|^2 =1\;.
\ee
The respective 3-dimensional group is denoted by M{\"o}b($\mathbb{D}$). The conformal boundary $\partial \mathbb{D}$ is invariant under transformations from M{\"o}b($\mathbb{D}$).

Geodesic lines on the Poincare disk are segments of circles orthogonal  to the (conformal) boundary of the disk. Circles on the complex plane are described by the equation
\be
\label{circle}
\gamma\, \bar{z}z+\beta z+\bar{\beta}\bar{z}+\gamma=0\;,
\qquad \gamma\in \mathbb{R}\;,
\quad  \beta\in \mathbb{C}\;.
\ee
At $\gamma = 0$ we find the diameters. A geodesic segment between points $z_1, z_2 \in \mathbb{D}$ can be mapped by an element of M{\"o}b($\mathbb{D}$) to an interval $(0,u)$ on the diameter. Then, its length can be easily calculated to be
\begin{equation}
\label{length}
L_{\mathbb{D}}(z_1, z_2)= \log \frac{1+u}{1-u} \;,
\qquad \text{where} \qquad   u=\left|\frac{z_2-z_1}{1-\bar{z}_1z_2}\right| \;.
\end{equation}
Indeed, there always exists a transformation $\in$  M{\"o}b($\mathbb{D}$)  that maps a given geodesic to a diameter such that a distinguished point on the geodesic goes to the center of $\mathbb{D}$. In our case this is \eqref{mob} with $a=1$ and $b=-z_1$ and the distinguished  point is $z_1$.

\subsection{Holographic Steiner trees}
\label{sec:holographic}

\begin{figure}[H]
  \centering
  \begin{minipage}[h]{0.3\linewidth}
    \includegraphics[width=1\linewidth]{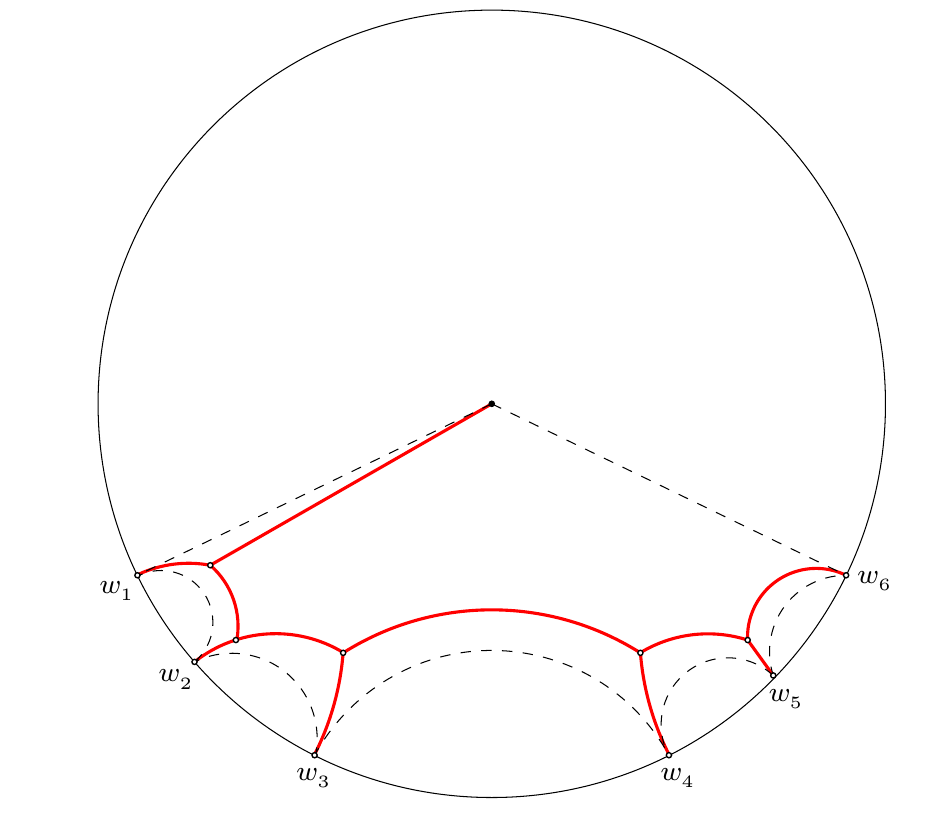}
  \end{minipage}
  \qquad\qquad
   \centering
  \begin{minipage}[h]{0.3\linewidth}
    \includegraphics[width=1\linewidth]{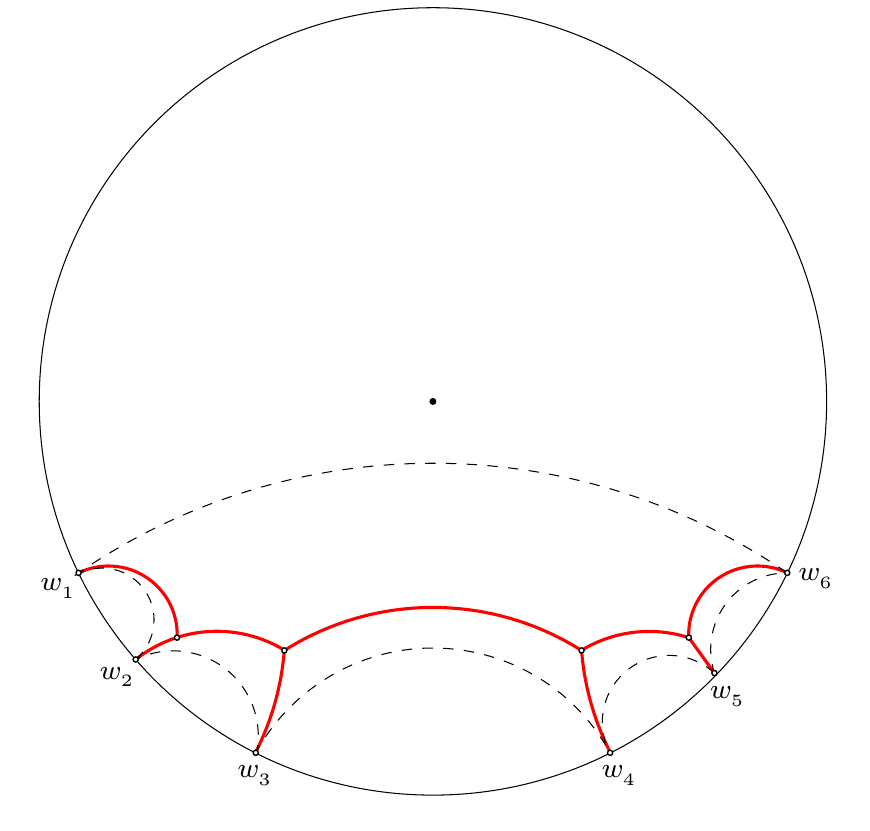}
  \end{minipage}
\caption{Holographic Steiner trees: (a) non-ideal $N=7$ graph, one edge ends in the center, (b) ideal $N=6$ graph, all edges end on the boundary circle. The associated hyperbolic polygons are shown in dashed lines. }
\label{Steiner_block}
\end{figure}

Steiner trees on $\disk$ dual to conformal blocks have a particular form, see Fig. \bref{Steiner_block}. We call them {\it holographic} Steiner trees. For a given $n$-point conformal block the holographic Steiner tree has $N = n-1$ endpoints, $N-1$ of which lie on  the  boundary $\bnd$ and one endpoint is the center of the disk. There are $N_F = N-2$ trivalent vertices which are the Fermat-Torricelli points. In total, the corresponding Steiner tree has $2N-2$ points connected by $2N-3$ edges among which there are $N$ {\it outer} edges and $N-3$ {\it inner} edges (also called {\it bridges} in graph theory).
\vspace{-2mm}

All points of a given Steiner tree including the endpoints and vertices are not invariant with respect to general conformal transformations from M{\"o}b($\mathbb{D}$) \eqref{mob}. For example, the center of the disk can be shifted away from its position. On the other hand, boundary endpoints always remain on the boundary because the boundary circle is invariant with respect to M{\"o}b($\mathbb{D}$).  Therefore, it is admissible to have a holographic Steiner tree with an endpoint not in the center of the disk. Such a tree is conformally equivalent to that one with an endpoint in the center.\footnote{From the boundary CFT perspective this is quite natural because two heavy background fields are coupled to the exchange channel in a point which is not specified, being instead an integration variable.}

Any Steiner tree can be inscribed in a hyperbolic polygon with $N$ corners. In our case, $N-1$ corners necessarily lie on the boundary circle and, therefore, the angles are zero. In hyperbolic geometry such corners are called {\it ideal}.  This property defines holographic Steiner trees: such trees can be inscribed into an $N$-gon with $N-1$ ideal vertices. Note that the number of ideal vertices is a conformal invariant.

%

The total length formula \eqref{minimum} for the holographic Steiner tree on the Poincare disk can  be  represented  as the sum over inner and outer edges,
\be
\label{LN}
L^{(N)}_{\mathbb{D}}(w) = \sum_{i\in outer} \epsilon_i\, L_{\mathbb{D}}(z_i, x_{i+1}) + \sum_{j\in inner} \tepsilon_j\, L_{\mathbb{D}}(x_j, x_{j+1})\;,
\ee
where all weights are divided into two subsets of outer weights $\epsilon_i$, $i = 0,...,n-2$ and inner weights $\tepsilon_{j}$, $j = 1,..., n-4$, the center $z_0 =0$,  the boundary endpoints have coordinates $z_j = \exp[iw_j]$, $j = 1,..., n-2$, the FT points have  coordinates $x_i$, $i = 1,..., n-3$.  A length $L_{\mathbb{D}}$ of each edge is given by the general formula \eqref{length}, and, therefore, to calculate  the total length we need to know explicit positions of the FT points. Note that the FT points are completely defined in terms of the endpoint positions and the edge weights by virtue of the minimal length condition. In order to find the length of the same Steiner tree on the Poincare disk with the angle deficit we have  to rescale the arguments according to
\be
\label{L_rescaled}
L^{(N)}_{\wdisk} := L^{(N)}_{\mathbb{D}}(\alpha w)\;,
\ee
see Section \bref{sec:disk}.
\vspace{-2mm}
\begin{figure}[H]
\centering
\includegraphics[width=55mm]{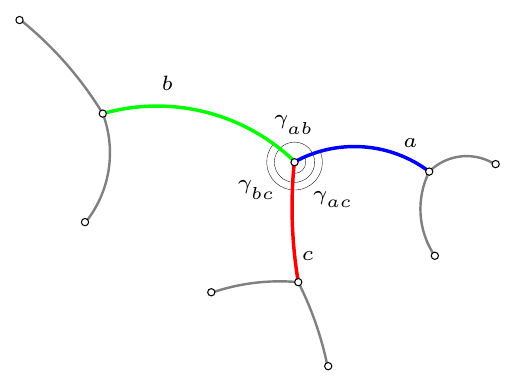}
\caption{Steiner tree on the non-Euclidean  plane with four Fermat-Torricelli  points. The middle  trivalent vertex is highlighted such that the edges of different weights are shown in different colors. The angles between edges are defined according to \eqref{cooos}.}
\label{vertex}
\end{figure}

Let us describe the analytic geometry method of finding the FT points. To this end, we note that the slope coefficient in the geodesic equation \eqref{circle} is given by
\be
\label{slope}
\varkappa = i\,\frac{\beta + \bar \beta + \gamma(z+\bar z)}{\beta -\bar \beta -\gamma(z-\bar z)}\;.
\ee
Consider now two edges $a$ and $b$ intersecting at angle $\gamma_{ab}$. Each edge is described by the geodesic equation \eqref{circle} so that using their slope coefficients \eqref{slope} we can find the angle of intersection
\be
\label{tangent}
\cos \gamma_{ab}   = \frac{1+ \varkappa_a \varkappa_b}{\sqrt{(1+\varkappa_a^2)(1+\varkappa_b^2)}}\;,
\ee
where $\varkappa_a$ and $\varkappa_b$ are the slopes of the tangent lines of the geodesic segments with weights $\epsilon_a$ and $\epsilon_b$ evaluated in a given FT point $z = x_i$. On the other hand, the angles are completely fixed by the edge weights \eqref{cooos}, and, therefore, we find the  relation
\be
\label{slope_eq}
\left(1+ \varkappa_a \varkappa_b\right)^2  -  \left(\frac{\epsilon^2_a-\epsilon^2_c - \epsilon^2_b}{2 \epsilon_a \epsilon_c}\right)^2 (1+\varkappa_a^2)(1+\varkappa_b^2)=0\;,
\ee
which is the second order polynomial equation in $\varkappa_a$ and $\varkappa_b$. Writing down the analogous equations for $b,c$ and $c,a$ we obtain the equation system for all slopes at given FT point. The system contains  three equations, however, the angles sum up to $2\pi$, and, thus, there are just two independent equations sufficient to find the slopes as functions of weights, $\varkappa_a = \varkappa_a(\epsilon, \tepsilon)$.

On the other hand, a given edge is described by the geodesic equation \eqref{circle} where coefficients depend on  endpoints $z_i$ and/or the FT points $x_j$. It follows that generally we have coordinate dependent slopes $\varkappa_a = \varkappa_a(z,x)$. Therefore, there are equations of the type $\varkappa_a(z,x) = \varkappa_a(\epsilon, \tepsilon)$ that allows  to find the FT coordinates explicitly in terms of the edge weights and the endpoint coordinates. The complexity of calculations grows rapidly with the number of the FT points $N_F$. Indeed, we have at least $2N_F$ equations with radicals which encode the slope and the cosine equations discussed above. Already in the $N_F = 1$ case we are confronted with a 4th order equation. In Section \bref{sec:examples} we  calculate all possible holographic Steiner trees with a single  FT point and a particular Steiner tree with two FT points.

\subsection{Cuts and connectivity }
\label{sec:cut}

Let $G_N$ denote a holographic Steiner tree with $N$ endpoints.  By definition, it is a connected graph because there is a path between any pair of vertices and/or endpoints.  However, removing an edge  may render the graph disconnected. Since there are two types of edges, inner and outer, we conclude that removing an outer edge does not disconnect the graph, while cutting an inner edge disconnects the graph in two connected subgraphs.

Let us consider a vertex formed by three edges labelled $a,b,c$ with  weights $\epsilon_{a,b,c}$, see e.g. Fig. \bref{vertex}. Cutting an edge is tantamount to that its weight is set to zero. Let  $\epsilon_a = 0$. Since the weights satisfy the triangle inequalities \eqref{triangle} we conclude that other two weights become equal, $\epsilon_b = \epsilon_c$. According to \eqref{cooos} the angle between the remaining edges become $\gamma_{bc} = \pi$ and, therefore, they merge into a single edge of a fixed weight $\epsilon_b$.

Suppose now that we remove an outer edge. Following the discussion above, the corresponding vertex vanishes and we have one less outer edge. The resulting graph remains  connected but with a smaller number of endpoints, i.e.
\be
\label{cut1}
G_N \hookrightarrow G_{N-1}\;,
\ee
where the hooked arrow denotes an outer edge cut. On the other hand, cutting an inner edge saves the number of endpoints, but splits the original graph into two independent subgraphs, each with less endpoints,
\be
\label{cut2}
G_{N} \leadsto G_{N_1} \cup G_{N_2}\;,
\qquad N_1+N_2 = N\;,
\ee
where the wavy arrow denotes an inner edge cut.

Holographic Steiner trees with outer edges  attached to the boundary $\partial \disk$ only will be  called {\it ideal}. These graphs can be inscribed into an ideal hyperbolic $N$-gon.  Respectively, {\it non-ideal} trees have an outer edge attached to the center of the disk.   An ideal holographic Steiner tree can be obtained from a non-ideal one by an edge cutting \eqref{cut1} or \eqref{cut2}, see Fig. \bref{Steiner_block}. Finally, we note that for a given $N$ there are two holographic Steiner trees, ideal and non-ideal ones.

\begin{figure}[H]
\centering
\includegraphics[width=105mm]{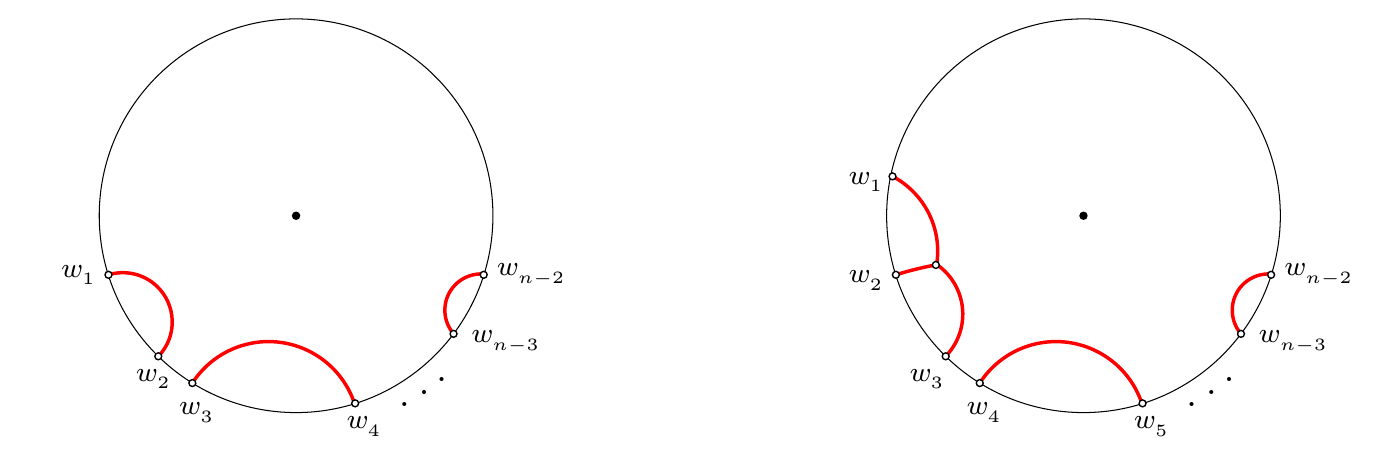}
\caption{Examples of disconnected holographic Steiner trees obtained from the Steiner tree $G_N$ by a maximal number of inner edge cuts using \eqref{cut2}. The resulting graphs are ideal Steiner trees. }
\label{dis}
\end{figure}

From the boundary CFT  perspective, ideal Steiner trees correspond to identity blocks and the edge cuts \eqref{cut1} and \eqref{cut2} are reformulated as factorization relations for conformal blocks, see Section \bref{sec:factor}.

\section{Examples of lower $N$ Steiner trees }
\label{sec:examples}

In what follows we apply the general procedure of finding the FT points described in the previous section to the case of holographic Steiner trees with a small number of endpoints $N=2,3,4$. These Steiner trees have at most two FT points and dual to particular  $3,4,5,6$-point conformal blocks (see Section \bref{sec:lower}).

\subsection{$N=2$ holographic Steiner trees}

\begin{figure}[H]

\centering
\includegraphics[width=85mm]{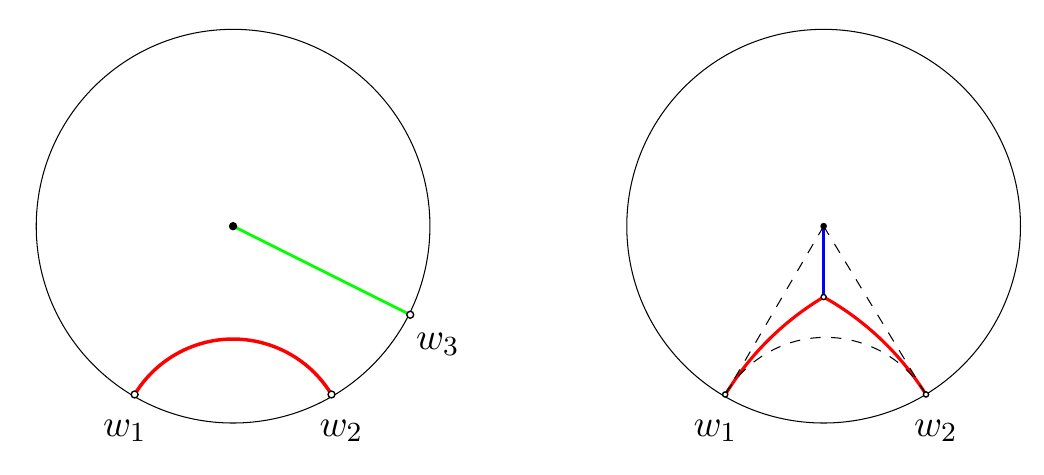}
\caption{(a) Two types of $N=2$ holographic Steiner trees. (b) A hyperbolic isosceles triangle (3-gon, dashed lines) with two ideal vertices and inscribed $N=3$ Steiner tree. Different colors imply different weights.  This $N=3$ tree can be viewed as the deformation of the $N=2$ three on the left, see the discussion around  \eqref{cos1}.}
\label{N2}
\end{figure}

There are two $N=2$ holographic Steiner trees both with  just one edge, see (a) Fig. \bref{N2}. The total length \eqref{minimum} in this case  is given  by \eqref{length}. However, we are faced here with the general  problem  that if at least one of endpoints lies on the boundary circle then the resulting length is infinite and must be regularized somehow.

Let us consider first the $N=2$ holographic tree with two boundary endpoints
$z_1=\exp[i w_1]$ and $z_2=\exp[i w_2]$. To regularize the length function we shift the points inside the disk as $z_1 = \exp[-\varepsilon + i w_1]$ and $z_2=\exp[-\varepsilon+i w_2]$ at $\varepsilon \to +0$. In Appendix \bref{app:A} we find that the length function \eqref{length} can be expanded with respect to $\varepsilon$ as
\be
\label{reg_length}
L^{(2)}_{\mathbb{D}}(w_2, w_1) = \log\left[4\sin^2\frac{w_2 - w_1}{2}\right] - 2\log \varepsilon + \cO(\epsilon)= 2\log\sin\frac{w_2 - w_1}{2}+\log 4- 2\log \varepsilon + \cO(\epsilon) \;.
\ee
Usually, from the holographic duality perspective, only the first term is relevant because we keep track of $w$-dependent terms  and do not care about (in)finite constant contributions. However, we define the regularized length as all $w$-dependent and $\varepsilon$-independent terms modulo constants. In \eqref{reg_length} this is the first term only. Such a definition gives rise to a negative length because we discarded $\epsilon$-dependent and constant terms that were making the original (non-regularized) length function \eqref{length} positive. Finally, we postulate the rescaled length \eqref{L_rescaled} to be
\be
\label{arc}
L^{(2)}_{\wdisk}  = 2\log\sin\frac{\alpha(w_2 - w_1)}{2}\;.
\ee

The $N=2$ holographic tree with one boundary endpoint can be considered along the same lines, see Appendix \bref{app:A}. Using \eqref{regL} we find that in this case the regularized length is zero\be
\label{radial}
L^{(2)}_{\wdisk} =0\;.
\ee

Now, let us shortly discuss what happens if we add an outer edge ending in the center of the disk, see (b) on Fig. \bref{N2}. Choosing $\epsilon_a = \epsilon_b \equiv \epsilon$ (red) and $\epsilon_c \equiv \tilde \epsilon$ (blue) we find from the cosine formula \eqref{cooos} that $\gamma_{ac} = \gamma_{bc}$,
and
\be
\label{cos1}
\cos\gamma_{ab} =  -1 +  \frac{\tilde \epsilon^2 }{2\epsilon^2}\;,
\qquad
\cos\gamma_{ac} =  - \frac{\tilde \epsilon }{2 \epsilon}\;.
\ee
We see that the ratio $\tilde\epsilon/\epsilon$ measures the deviation from the $N=2$ graph. The angles are   $\gamma_{ab}  \leq  180^{\circ}$  and $\gamma_{ac}=\gamma_{bc}  \geq  90^{\circ}$. Pictorially, it corresponds to pulling the Fermat-Torricelli point towards the center of the disk along the outer edge (blue). When $\tepsilon = 0$ the associated hyperbolic triangle degenerates and the FT point lies down on the arc.

\subsection{$N=3$ holographic Steiner trees}
\label{sec:N3}
\begin{figure}[H]
\centering
\includegraphics[width=130mm]{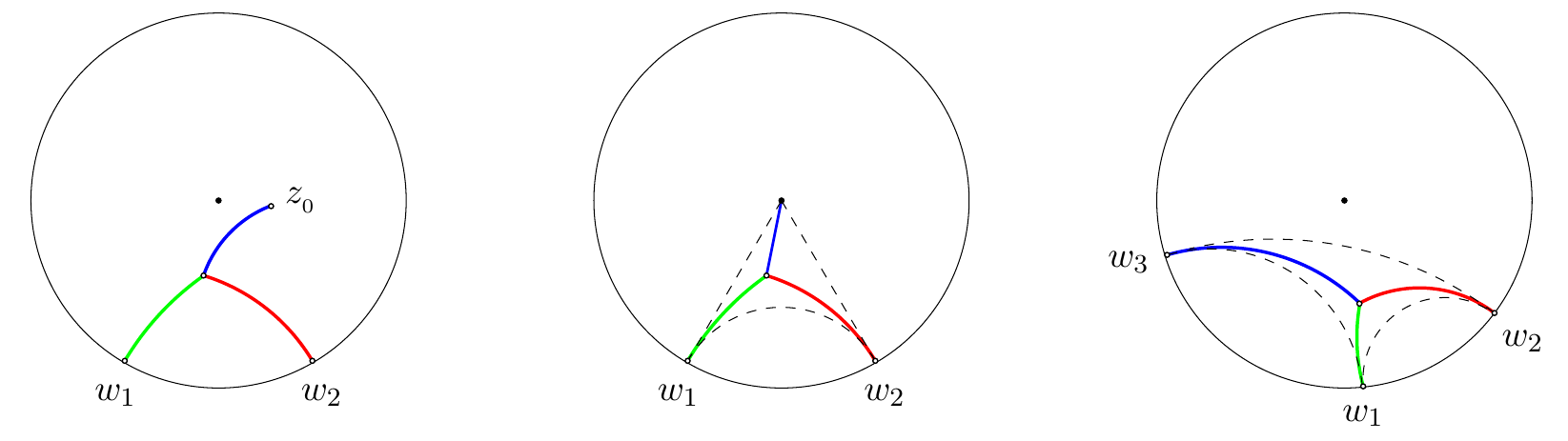}
\caption{(a) Master $N=3$  Steiner tree with two boundary endpoints and an arbitrary third endpoint $z_0 \in \disk$, (b) $z_0=0$ gives the holographic non-ideal $N=3$ Steiner tree dual to the general 4-point conformal block (c)  $z_0\in \bnd$ gives the holographic ideal $N=3$ Steiner tree dual to particular  5-point identity conformal block. The associated  hyperbolic 3-gons are shown in dashed lines.}
\label{N3pic}
\end{figure}

Let us note that all $N = 3$ holographic Steiner trees can be generated from a single $N=3$ tree with two boundary endpoints $z_{1,2}\in \bnd$ and an arbitrary endpoint $z_0\in \disk$, see (a) on Fig. \bref{N3pic}. Edges have arbitrary weights $\epsilon_{1,2}$ and $\epsilon_0$. The  length function $L^{(3)}_\disk$ of the {\it master} tree is parameterized by $z_0$. Sending $z_0$ either to the center of the disk ($z_0=0$) or the boundary ($|z_0| = 1$) we find lengths of two other Steiner trees, see (b) and (c) on Fig. \bref{N3pic}. These last two graphs exhaust all possible $N=3$ holographic Steiner trees (ideal and non-ideal Steiner trees, see Section \bref{sec:cut}).

Let us find the length function of the graph (a) on \bref{N3pic}. To this end, we explicitly write down and solve the equation system \eqref{slope_eq}, see Appendix \bref{app:N34} for detailed calculations. We represent coordinates of all endpoints $z_{0,1,2}$ as $z_{1,2}=\exp[iw_{1,2}]$ and $z_{0} = r_0\exp[i w_0]$ and introduce new convenient parameterization
\be
\label{new}
\ba{c}
\dps \gamma = \frac{\epsilon_1 + \epsilon_2}{\epsilon_0}\;,
\quad
\beta = \frac{\epsilon_1 - \epsilon_2}{\epsilon_0}\;,
\qquad
\dps |\gamma|\geq 1 \;, \quad |\beta|\leq 1\;;
\qquad
w_{ij} = \frac{w_i-w_j}{2}\;.
\ea
\ee

Now, we introduce auxiliary functions
\be
\label{final_solution_t}
\ba{l}
\dps L_{0} = \frac{\sqrt{\gamma ^2-1} }{\sqrt{1-\beta ^2} (\gamma +1)}\left(\sqrt{P-1}-\sqrt{P-\beta^2}\right)\;,
\vspace{-2mm}

\\
\\
\dps L_{1} = \frac{(\beta +\gamma) \sqrt{1-\beta^2}}{\sqrt{\gamma^2 - 1}} \frac{\left(\sqrt{P-\beta ^2 }+\beta  \sqrt{P-1}\right)}{K_2}\sin^2 w_{21}\;,
\vspace{-2mm}

\\
\\
\dps L_{2} = \frac{ (\gamma -\beta)  }{\sqrt{\gamma ^2 - 1}\sqrt{1-\beta^2}} \frac{K_2}{\left(\sqrt{P-\beta ^2 }+\beta  \sqrt{P-1}\right)}\;,
\ea
\ee
where
\be
K_{1,2} = \frac{1+r_0^2 - 2r_0 \cos(w_{1,2}-w_0)}{1-r_0^2}\;,
\qquad
P = \frac{K_1 K_2}{\sin^2  w_{21}}\;.
\ee
Then, the length function of the master $N=3$ Steiner tree is given by
\be
\label{masterN3}
L^{(3)}_{\mathbb{D}}(w_1,w_2,z_0) =  \epsilon_0 \log L_0 + \epsilon_1\log L_1 +\epsilon_2\log L_2\;.
\ee
Note that arguments $w_1$ and $w_2$ enter the length function only through the combination  $w_{21} = (w_2-w_1)/2$ which is the semiangle position of two boundary endpoints.

Let us now find the lengths of two other graphs (b) and (c) on Fig. \bref{N3pic}. These are $N=3$ holographic Steiner trees of interest.

\paragraph{$z_0=0$ case.}  From \eqref{masterN3} we find the length function\footnote{The $w$-dependent part of this formula  relevant for the holographic duality analysis was obtained using the worldline formalism in a different parametrization of the bulk space \cite{Hijano:2015rla}. For equal weights $\beta=0$ formula \eqref{general_4pt_length} was found in \cite{Alkalaev:2015wia}.}
\be
\label{general_4pt_length}
\ba{l}
\dps
L^{(3)}_{\disk}(w_1,w_2) = \epsilon_0\,\text{Arctanh}\, \dps \frac{\cos w_{21}}{\sqrt{1-\beta^2 \sin^2 w_{21}}} +
\\
\\
\hspace{22mm}+ \dps\epsilon_0 \left[\gamma \log \sin w_{21} - \beta  \log \left(\beta \cos w_{21} + \sqrt{1 - \beta^2 \sin^2 w_{21}}\right)\right] + C_0\;,
\ea
\ee
where we introduced the $w$-independent function of the weights,
\be
\label{C0}
 C_0 = \frac{ \epsilon_0}{2}\left(\log \frac{\gamma-1}{(\gamma+1)(1-\beta^2)} + \gamma \log \frac{\gamma^2-\beta^2}{(\gamma^2-1)(1-\beta^2)} +\beta \log  \frac{\gamma+\beta}{(1-\beta^2)(\gamma-\beta)}\right)\;.
\ee

\paragraph{$|z_0|=1$ case.} Evaluating the length function at $z_0$ on the boundary is more tricky because the length function diverges so we have to use the $\varepsilon$-prescription. To this end, we shift the radial position of the third point  $r_0=1-\varepsilon$ and denote its argument as $w_0 \equiv w_3$ and weight as $\epsilon_0 \equiv \epsilon_3$. Then, in the limit $\varepsilon\rightarrow+0$ we have $K_{1,2}=2\varepsilon^{-1} \sin^2\frac{w_3-w_{1,2}}{2} + O(\varepsilon)$, and keeping the leading term only (see Appendix \bref{app:A}) we obtain from \eqref{masterN3} the limiting length function
\be
\label{vacuuma_5pt_length}
\ba{c}
\dps L^{(3)}_{\disk}(w_1,w_2,w_3) =  \epsilon_1\log \frac{\sin w_{21}\sin w_{31}}{\sin w_{32}} +  \epsilon_2\log \frac{\sin w_{21}\sin w_{32}}{\sin w_{31}} + \epsilon_3\log \frac{\sin w_{31}\sin w_{32}}{\sin w_{21}}
+C_1\;,
\ea
\ee
where, taking into account \eqref{C0} we introduced the $w$-independent function (constant) of the weights,
\be
\label{C1}
C_1 = C_0 + \epsilon_3\Big[\beta\log(1+\beta) - \frac{1}{2}\log (1-\beta^2)\Big]\;.
\ee

\subsection{$N=4$ holographic Steiner trees}
\label{sec:N4}

\begin{figure}[H]
\centering
\includegraphics[width=100mm]{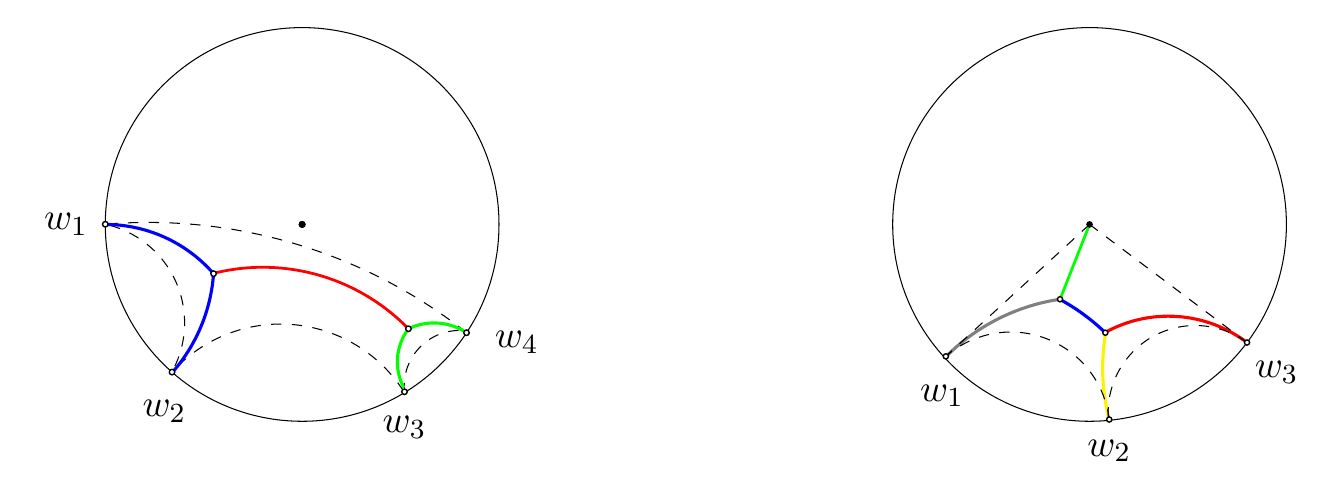}
\caption{Two types of $N=4$ holographic Steiner trees: (a) ideal holographic Steiner tree, this is the hyperbolic analog of the graph (b) on Fig. \bref{euclid}, (b) non-ideal holographic Steiner tree. Cutting the bridge (blue inner edge) yields the disconnected graph (a) on Fig. \bref{N2}, see Section \bref{sec:cut}. The associated  hyperbolic 4-gons are shown in dashed lines.}
\label{N4}
\end{figure}

There are two types of $N=4$ holographic Steiner trees, Fig. \bref{N4}. In what follows we discuss only the first graph which corresponds to the identity 6-point conformal block, see Section \bref{sec:lower}. The second graph is dual to 5-point conformal block and will be considered elsewhere.\footnote{Both the 5-point  classical block and dual graph were extensively discussed  using various perturbative approximations in dimensions $\epsilon, \tepsilon$ \cite{Alkalaev:2015wia,Alkalaev:2015lca,Alkalaev:2015fbw,Datta:2014zpa,Belavin:2017atm}.    }

There are two FT points in the $N=4$ case. To find the total length $L_{\disk}^{(4)}(w_1, w_2, w_3, w_4)$ we may follow the general strategy described in the end of Section \bref{sec:holographic} and formulate the system of equations that can be solved to fix the FT points. However, we take a different route   and  view our $N=4$ tree as two $N=3$ trees glued together in some point. Using \eqref{masterN3} we  can write down
\be
\label{junction}
L_{\disk}^{(4)}(w_1, w_2, w_3, w_4) = L^{(3)}_{\mathbb{D}}(w_1,w_2,z_0) +L^{(3)}_{\mathbb{D}}(w_3,w_4,z_0)\;,
\ee
where the junction point $z_0$ is fixed by the minimization  condition \eqref{minimum}. This procedure is explicitly described in Appendix \bref{app:N34}.

To simplify the calculations we consider equal weights $\epsilon_{1}= \epsilon_{2}$, $\epsilon_{3}=\epsilon_{4}$, and $\epsilon_{1}\neq \epsilon_{3}\neq \tilde\epsilon$. In this case we obtain
\be
\label{answ_br}
\ba{c}
\dps L^{(4)}_{\disk}(w_1,w_2,w_3,w_4) = 2\epsilon_{1}\log\sin w_{21} + 2\epsilon_{3}\log\sin w_{43} - 2\tilde \epsilon \log \left(\sqrt{U+1} -\sqrt{U}\right)\;,
\ea
\ee
where $U$ is the trigonometric ratio
\be
\label{defU}
\dps U = \frac{\sin w_{32}\sin w_{41}}{\sin w_{43}\sin w_{21}}\;.
\ee
Here, the first two terms are given by length functions  \eqref{reg_length}, while the last term is the length of the bridge, we also omitted the constant term $\tilde C$, see \eqref{B13}.   Note that independent variables are in fact $\sin w_{ij}$.

\subsection{Fermat-Torricelli point via M{\"o}bius  transformation}

As we have seen, finding positions of  FT points is a complicated problem that amounts to solving polynomial equations. In this section we give an example how to calculate FT points using the M{\"o}bius  transformations of the disk. We shall use two basic properties of M{\"o}b($\mathbb{D}$): (a) the boundary $\partial \disk$ is invariant, (b)  angles are preserved.

  Let us consider the $N=3$ Steiner tree with three points on the boundary, see (c) on Fig. \bref{N3pic}. In this case there is only one FT point and its position is determined by  boundary endpoints $w_i$ and weights $\epsilon_{i}$, where $i=1,2,3$.

\begin{figure}[H]
\centering
\includegraphics[width=100mm]{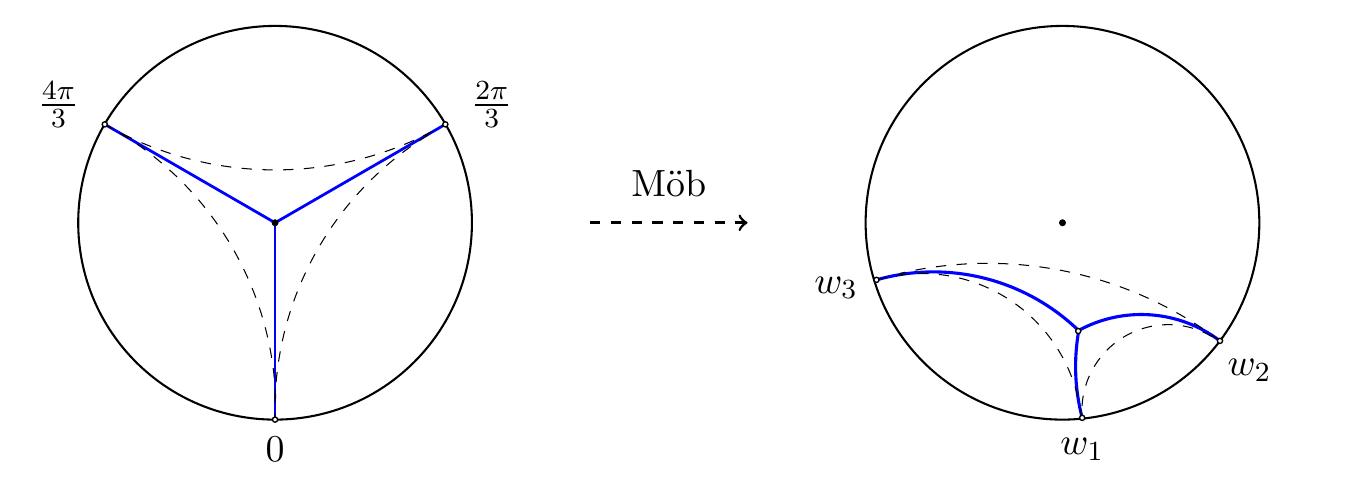}
\caption{The holographic ideal N = 3 Steiner tree (c) on Fig. \bref{N3pic} can be obtained by the M{\"o}bius  transformation from the graph with the FT point in the center of the disk. The weights are equal and the angles at both FT points are equal $120^{\circ}$.  }
\label{trans}
\end{figure}

Now consider the simplest version of such  $N=3$ Steiner tree when the FT point  is at the center of the disk. In this case, choosing $w_1=0$ we find that the endpoints are uniquely fixed by the cosine formulas \eqref{cooos}. E.g. for equal weights $\epsilon_1 = \epsilon_2 = \epsilon_3$ we have $w_1=0$, $w_2 =2\pi/3$, $w_3 =4\pi/3$. Then, one can use a three-parametric M{\"o}bius  transformation \eqref{mob} that takes boundary endpoints $(0,2\pi/3, 4\pi/3)$ to  boundary endpoints $(0,w_2,w_3)$. Such a transformation translates the simplest graph into the general $N=3$ graph with three  boundary endpoints (see Fig. \bref{trans}). The parameters  of the  transformation are completely determined by specifying  initial three and final three endpoints,
\be
\label{ctpc}
\ba{c}
\dps z \to z^\prime = \frac{a z + b}{\bar b z +\bar{a}}\;,
\\
\\
\dps a =  \frac{\exp\left[\frac{i w_{_2} - i w_{_0} + i w_{_3}}{2}\right] - \exp\left[\frac{i w_{_2} + i w_{_0} - i w_{_3}}{2}\right]  + \exp\left[\frac{i w_{_2} + i w_{_0} + i w_{_3}}{2}\right] - \exp\left[\frac{i w_{_3} - i w_{_0} - i w_{_2}}{2}\right]}{2\sqrt{(\sin w_2 + \sin(w_3 - w_2) - \sin w_3)\sin w_0}}\;,
\\
\\
\dps b = \frac{ \exp\left[\frac{i w_{_2} - i w_{_3} - i w_{_0}}{2}\right] + \exp\left[\frac{i w_{_0} + i w_{_3} - i w_{_2}}{2}\right] - \exp\left[\frac{i w_{_2} + i w_{_3} + i w_{_0}}{2}\right] - \exp\left[\frac{i w_{_2} + i w_{_3} - i w_{_0}}{2}\right]}{2\sqrt{(\sin w_2 + \sin(w_3 - w_2) - \sin w_3)\sin w_0}}\;,
\ea
\ee
where we denoted $w_0 = 2\pi/3$. The FT point flows from the center to some new point
\be
\label{z_ft}
z_{_{FT}} = \frac{\exp[iw_2 + iw_3]-\exp[i w_2] + \exp[iw_2 + iw_3 + iw_0] - \exp[iw_3 + i w_0]}{\exp[i w_2 + i w_0] + \exp[i w_3] - \exp[i w_0] - 1 }\;.
\ee

Further, using \eqref{z_ft} and \eqref{regL} one can calculate lengths of three edges of the resulting graph. In this way we obtain the total length of an arbitrary  ideal $N=3$ Steiner tree and the final expression coincides with \eqref{vacuuma_5pt_length}. One can also generalize to the case of arbitrary weights $\epsilon_1 \neq \epsilon_2 \neq \epsilon_3$.

\section{Perturbative classical conformal blocks}
\label{sec:perturbative}

\begin{figure}[H]
\centering
\includegraphics[width=120mm]{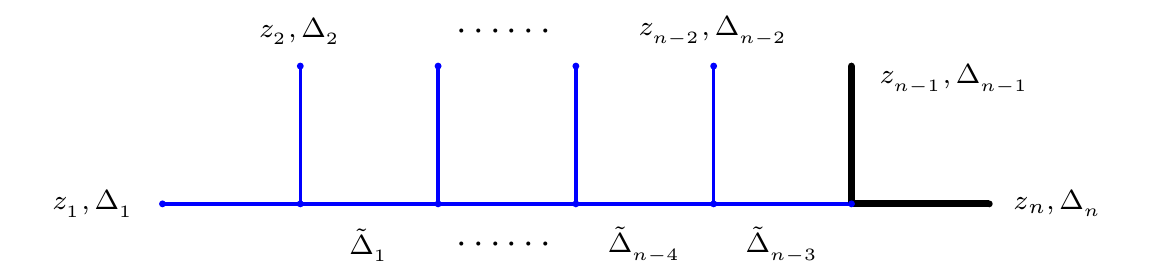}
\caption{The $n$-point conformal block with two heavy background primary operators depicted by two bold black lines.}
\label{block}
\end{figure}

In this and subsequent sections we discuss perturbative large-$c$ regime of two-dimensional CFT.  We do not consider  correlation functions focusing instead on conformal blocks. This is a crucial simplification since we completely ignore CFT data. On the other hand, conformal blocks are still interesting to consider because these functions form a  basis in the space of correlators. In what follows we choose a particular OPE channel  which generalizes the $s$-channel of 4-point correlation functions to the higher-point case (this is the comb diagram on Fig. \bref{block}).

Let $\cF_n(z| \Delta, \tilde \Delta,c)$ be a holomorphic conformal block of the $n$-point correlation function of $n$ primary operators in points  $z = \{z_1,...\,,z_n\}$ on the complex plane with (holomorphic) conformal dimensions $\Delta = \{\Delta_1, ...\Delta_n\}$ and exchange channel dimensions $\tilde \Delta = \{\tilde\Delta_1, ..., \tilde \Delta_{n-3}\}$, the central charge is $c$ \cite{Belavin:1984vu}.  Suppose now that  conformal dimensions $\Delta$ and $\tilde \Delta$ depend linearly on the central charge, i.e. these are {\it heavy} dimensions  $\Delta, \tilde \Delta = \cO(c^1)$.  Then, decomposing $\cF_n(z| \Delta, \tilde \Delta,c)$ near $c=\infty$ we find out that the block function can be represented in the exponentiated form \cite{Zamolodchikov:1987ie}
\be
\label{class}
\cF_n(z| \Delta, \tilde \Delta,c) \,\Big |_{c\to\infty} \;\rightarrow\; \exp\big[\,\frac{c}{6}f_n(z| \epsilon, \tilde \epsilon)\,\big]\;,
\ee
where the exponential factor $f_n(z| \epsilon, \tilde \epsilon)$ is the $n$-point {\it classical} conformal block which depends on external and intermediate  classical conformal dimensions $\epsilon = \{\epsilon_1, ..., \epsilon_n\}$ and $\tilde \epsilon = \{\tilde\epsilon_1,..., \tilde\epsilon_{n-3}\}$
\be
\epsilon_i = \frac{6\Delta_i}{c}\;,
\qquad
\tepsilon_j = \frac{6\tilde\Delta_j}{c}\;.
\ee
In general, already 4-point classical blocks are quite complicated functions yet unknown in the closed form. In what follows, we calculate conformal blocks using the heavy-light perturbation theory in conformal dimensions \cite{Fitzpatrick:2014vua} (see also \cite{Hijano:2015rla,Alkalaev:2015wia,Banerjee:2016qca,Belavin:2017atm}).

\subsection{Heavy-light approximation}
\label{sec:HL}

Let us consider the large-$c$ regime and all conformal dimensions are heavy. Suppose now that two of external primary heavy operators with equal dimensions $\Delta_{n-1} = \Delta_n = \Delta_h$ are much heavier than other primary operators and  exchange channels, i.e.
\be
\Delta_i/\Delta_h \ll 1\;,
\qquad
\tilde\Delta_j/\Delta_h \ll 1\;,
\ee
at $i=1,..., n-2\;, j = 1,..., n-3$.

It means that the respective classical block function  $f(z|\epsilon, \tilde\epsilon)$ can be represented by the Taylor series near the point $\epsilon_i,\tilde\epsilon_j =0$, $i=1,..., n-2$, $j = 1,..., n-3$ in the conformal parameter space. The leading term is obviously zero because the conformal block of 2-point function of the background operators vanishes identically, while the next-to-leading term is the {\it perturbative} classical block which we  denote by $f_n(z|\alpha,\epsilon, \tilde\epsilon)$, where $\alpha = \sqrt{1-24 \Delta_h/c }$.

\subsection{The block/length correspondence}

Here we formulate the semiclassical $\dual$ correspondence briefly discussed in Introduction (for  review and references see e.g. \cite{Alkalaev:2016rjl}). We consider CFT  on the boundary plane with two background heavy operators $\alpha = \sqrt{1-24 \Delta_h/c }$, where $\Delta_h < c/24$, and  interacting matter fields in the $\ads$ spacetime with an angle deficit $\wads$. The correspondence works in the large-$c$ regime when conformal dimensions are heavy $\Delta = \cO(c^1)$. Then, there are $n-2$ heavy primary operators which within the heavy-light approximation correspond to point particles with masses  $\sim \Delta/c$ propagating on the $\wads$ background.

The spacetime $\wads$ is the rigid cylinder in which we consider a constant time slice $\wdisk$ (see Section \bref{sec:disk}). Let $(z, \bar z)$ be coordinates on the punctured complex plane  and $(w, \bar w)$ be coordinates on the boundary cylinder  mapped to each other by the conformal transformation $w(z)  = i \ln (1-z)$. Let $G_{n-1}$ be a non-ideal holographic Steiner tree on the disk $\wdisk$ (see Sections \bref{sec:holographic} and \bref{sec:cut}). The correspondence between the perturbative  classical block and the holographic  Steiner tree reads
\be
\label{fS}
f_n(z(w)|\alpha,\epsilon, \tilde\epsilon) = -L^{(n-1)}_{\wdisk}(w|\epsilon, \tilde\epsilon)+ i\sum_{k=1}^{n-2}\epsilon_k w_k\;,
\ee
where the length function $L^{(n-1)}_{\wdisk}$ of $G_{n-1}$ is given by \eqref{L_rescaled}. A few comments are in order: (a) the weights and the classical dimensions are identified (note that we equate $\tilde \epsilon_{n-3}\equiv \epsilon_0$), (b) the classical conformal block is defined modulo constants so that all $w$-independent terms in $L^{(n-1)}_{\wdisk}$ are neglected, (c)  here $z,w \in S^1$, where the circle is realized as $\bnd_{\alpha}$ in the bulk and the unit circle on the boundary, $(z_i-1)(\bar z_i -1) = 1$ for $i=1,..., n-2$, (d) the fusion rules for the conformal block are now encoded in the triangle inequalities of the Steiner trees\footnote{The triangle inequalities \eqref{triangle} are analogous to triangle inequalities satisfied by conformal dimensions of primary operators in the semiclassical limit of the DOZZ three-point correlation function \cite{Harlow:2011ny}. Together with the Seiberg bound \cite{Seiberg:1990eb} and the Gauss-Bonnet constraint \cite{Harlow:2011ny} they guarantee the existence of real solutions in the Liouville theory. It would be interesting to derive the triangle inequalities  directly within the monodromy approach.} \eqref{triangle}, (e) the real part of the block is the length function, the imaginary part is given by $\epsilon_k w_k$ terms.

\subsection{Modified $n$-point monodromy equations}
\label{sec:modified}

The monodromy method for calculating classical conformal blocks is interesting because of its many conceptual and technical advantages (for review see e.g. \cite{Hartman:2013mia,Litvinov:2013sxa,deBoer:2014sna,Alkalaev:2016rjl}). For example, the monodromy method reformulated within the heavy-light perturbation theory is equivalent to the worldline approach in the bulk \cite{Hijano:2015rla,Alkalaev:2015lca,Alkalaev:2016rjl}. In this section we slightly modify the monodromy equations for the accessory parameters by relaxing the value of the first point $z_1$ which is usually fixed as  $z_1 =0$ by projective conformal transformations. Relaxing $z_1$ is required when considering  factorization relations for  identity conformal blocks in Section \bref{sec:factor}.

In order to find an $n$-point (perturbative) classical block one introduces $n$ auxiliary variables called accessory parameters $c_{i}$. These  are subjected to  $n$ algebraic equations, where the first three equations are linear,
\be
\label{3cons1}
\sum^n_{i = 1} c_i = 0\;,
\qquad
\sum^n_{i = 1} (c_i z_i + \epsilon_i) = 0\;,
\qquad
\sum^n_{i = 1} (c_i z_i^2 + 2\epsilon_i z_i) = 0\;.
\ee
It follows that the accessory parameters $c_2,...\,, c_{n-2}$ can be chosen  as independent. Indeed, using  the projective conformal invariance we can fix positions of  two  operators $z_{n-1} = 1, z_n = \infty$, while $z_1$ remains arbitrary. Then, fixing the heavy background dimensions as $\epsilon_{n-1} = \epsilon_n$ and we find solution to \eqref{3cons1} as
\begin{align}
&c_1 = -\frac{1}{1-z_1}\left[\sum^{n-2}_{i=2}\big[c_i(1-z_i)- \epsilon_i\big] -\epsilon_1\right],\label{ccc}\\
&c_{n-1} = -\frac{1}{1-z_1}\left[\sum_{i=2}^{n-2}c_i (z_i-z_1) +\sum_{i=1}^{n-2} \epsilon_i \right],
\qquad
c_n = 0\label{ccc1}\,.
\end{align}
The remaining $(n-3)$ equations for the accessory parameters within the heavy-light approximation take the form
\be
\label{moneq}
I_{+-}^{(n|k)}\,I_{-+}^{(n|k)}+\left(I_{++}^{(n|k)}\right)^2  + 4\pi^2 \tilde\epsilon^2_{k}=0\;,
\qquad k = 1,...\,, n-3\,,
\ee
where
$$
\ba{c}
\dps
I^{(n|k)}_{+-}  =\frac{2\pi i}{\alpha}\left[\left(1-z_1\right){}^{\alpha }\Big(\alpha \epsilon_1+\sum_{i=2}^{n-2}(c_i(1-z_i)-\epsilon_i)\Big)-\sum_{i=2}^{k+1}(1-z_i)^\alpha(c_i(1-z_i)-\epsilon_i(1+\alpha))\right],
\ea
$$
\be
\label{Is}
I^{(n|k)}_{-+}  = I^{(n|k)}_{+-}\big|_{\alpha \rightarrow -\alpha}\,,
\qquad\;\;
I^{(n|k)}_{++} = \frac{2\pi i}{\alpha}\sum_{i=k+2}^{n-2}\big[c_i (1-z_i)-\epsilon_i\big]\,.
\ee
In total, we have $n$ equations for $n$ variables and  the accessory parameters are particular roots
of the form $c_i = c_i(z|\alpha,\epsilon, \tilde\epsilon)$, where $z=(z_1,...\,,z_{n-2})$. Then, the perturbative classical block is defined by means of the following relations
\be
\label{acs1}
c_i = \frac{\partial}{\partial z_i}\,f_n(z|\alpha, \epsilon, \tilde\epsilon)\;,\qquad i = 1,...\,,n-2\;.
\ee
The system \eqref{acs1} can be solved  in the standard fashion by integrating in $z_1$ first  and isolating the integration constant which depends  on  $z_2,...\,,z_{n-2}$ only, then integrating in $z_2$, etc. Alternatively, noticing that the accessory parameters should satisfy the integrability condition $\partial_i c_j -\partial_j c_i= 0$ we may treat the equations \eqref{acs1} cohomologically and write down the formal solution (modulo integration constants) for any $n$ as
\be
\label{homotopy}
f_n(z|\alpha, \epsilon, \tilde\epsilon)  =  \int_0^1 dt\, z^i c_i(tz|\alpha,\epsilon, \tilde\epsilon)\;.
\ee
Since the accessory parameters typically diverge at $z\to 0$ as $1/z$ while  classical conformal blocks $f_n(z|\alpha, \epsilon, \tilde\epsilon) \to \log z$ at $z\to 0$ we conclude that the above integral is ill-defined at $t=0$ and must be regularized. This can be achieved if one redefines  classical conformal blocks by neglecting logarithmic terms that simply results in the standard  power-law prefactors $\sim z^\gamma$ for conformal block functions (it is a matter of different normalizations of conformal blocks). For classical blocks that are logarithms of the original block functions this means that redefined $f_n(z|\alpha, \epsilon, \tilde\epsilon) $ is regular near $z=0$ and \eqref{homotopy} is directly applicable. In fact, the homotopy formula captures the standard expansion of the $s$-channel type block near $z=0$.

\section{Lower point  perturbative blocks}
\label{sec:lower}

In this section, using the monodromy method we explicitly calculate  lower point perturbative classical blocks including  $4$-point general  block and  5,6-point identity blocks. Recall that when comparing with the Steiner trees by means of the formula \eqref{fS} we equate the classical dimension of the rightmost exchange   channel with the weight of the outer edge ending in the center of the disk, $ \tilde \epsilon_{n-3}\equiv \epsilon_0$. Also, it is convenient to introduce the following variables
\be
\label{P}
P_{i} = (1-z_i)^{\alpha} \;,\quad i =1,...,\,n-3\;.
\ee

\subsection{3-point general  block}
\label{sec:3pt}
Let us consider first the simplest case of 3-point blocks. The accessory parameters here are $c_1,c_2,c_3$ and the monodromy equations are reduced to the three linear conditions \eqref{3cons1} that are solved as
\be
c_1 = \epsilon_1 P_1^{1/\alpha}\;,
\qquad
c_2 = -c_1\;,
\qquad
c_3=0\;,
\ee
see \eqref{ccc}, \eqref{ccc1}, here we used the notation \eqref{P}. Then, according to \eqref{acs1} the block function is given by
\be
\label{3pt}
f_3(z_1|\epsilon_1) = \epsilon_1 \log P_1^{-1/\alpha}\;.
\ee
We note that the 3-point block does not depend on the background heavy dimension $\epsilon_h(\alpha)$, and, therefore, it is an exact result within the heavy-light approximation. In the bulk it corresponds to the  $N=2$ Steiner tree of the zeroth length \eqref{radial}, see (a) on Fig. \bref{N2}.   This is in agreement with the correspondence formula \eqref{fS}.

\subsection{4-point general  block}
\label{sec:4pt}
In this case, the monodromy equations \eqref{moneq} take the form
\be
\label{4pt_sys}
I^{(4|1)}_{+-}I^{(4|1)}_{-+} + 4\pi^{2}\tepsilon_1^{2} = 0\;.
\ee
This quadratic equation in $c_2$ can be directly solved\footnote{To the best of our knowledge both the accessory parameter and 4-point conformal block for arbitrary dimensions $\epsilon_{1,2}$ and $\epsilon_{0}$ were not given explicitly in the  literature. In the $\epsilon_1 =\epsilon_2$ case these expressions can be found in \cite{Fitzpatrick:2014vua,Hijano:2015rla}, while the conformal block with $\epsilon_1 \neq \epsilon_2 \neq \epsilon_0$ was given in \cite{Hijano:2015rla} within the bulk parametrization as the geodesic length. Moreover, in view of the factorization theorem of Section \bref{sec:factor} we represent the expression which depends on both points $z_1, z_2$ while usually $z_1=0$ ($P_1=1$).}

$$
\dps  c_1 = \epsilon_1 P_1^{-1/\alpha} - \alpha\frac{(\epsilon_1 + \epsilon_2)(P_1 +P_2) + \tilde\epsilon_1\sqrt{\beta^2 (P_1 - P_2)^2 + 4 P_1 P_2}}{2(P_1 - P_2)}\;,
$$

\be
c_2 = \epsilon_2 P_2^{-1/\alpha} + \alpha\frac{(\epsilon_1 + \epsilon_2)(P_1 +P_2) + \tilde\epsilon_1\sqrt{\beta^2 (P_1 - P_2)^2 + 4 P_1 P_2}}{2(P_1 - P_2)}\;,
\ee
where we have chosen just one root (in this case the conformal block has correct asymptotics, see the end of Section \bref{sec:modified}), and used parameterization \eqref{new}, $c_1$ is obtained from \eqref{ccc}. Integrating  equations \eqref{acs1} we find that up to $z$-independent terms the  conformal block is given by
\be
\ba{c}
f_{4}(z_{1,2}|\alpha,\epsilon_{1,2} ,\tilde\epsilon_{1}) = \epsilon_1 (-1 + \alpha)\log P_1^{1/\alpha} + \epsilon_2 (-1 + \alpha)\log P_2^{1/\alpha} - (\epsilon_1 + \epsilon_2)\log[P_1 - P_2] +
\\
\\
+ (\epsilon_1 - \epsilon_2)\log\left[\beta(P_1 + P_2) + \sqrt{4 P_1 P_2 + \beta^2 (P_1 -P_2)^2}\right] -
\\
\\
\dps -\frac{\tilde\epsilon_1}{2}\log\left[\frac{P_1 + P_2 +  \sqrt{4 P_1 P_2 + \beta^2 (P_1 -P_2)^2}}{P_1 + P_2 -  \sqrt{4 P_1 P_2 + \beta^2 (P_1 -P_2)^2}}\right]\;.
\ea
\ee
Now, using the correspondence formula \eqref{fS} we can explicitly see that
\be
f_{4}(z_{1,2}|\alpha,\epsilon_{1,2} ,\tepsilon_{1}) = - L^{(3)}_{\wdisk}(w_1,w_2|\epsilon_{0,1,2})  + i \epsilon_1 w_1 + i \epsilon_2 w_2\;,
\ee
where $L^{(3)}_{\wdisk}$ is given by $w$-dependent part of the length function of the holographic $N=3$ Steiner tree   \eqref{general_4pt_length}, see (b) on Fig. \bref{N3pic}.

In Section \bref{sec:factor}  we will need the so-called identity block obtained when the exchange channel has zero dimension, $\tilde\epsilon_1 = 0$. From the fusion conditions we get $\epsilon_1 = \epsilon_2$, and, therefore,
\be
\label{4vac_s}
f^{(1)}_{4}(z_{1,2}|\alpha,\epsilon_1) = \epsilon_1 ( \alpha - 1 )\left(\log P_1^{1/\alpha} + \log P_2^{1/\alpha}\right) - 2 \epsilon_1 \log[P_1 - P_2]\;.
\ee
\subsection{5-point identity  blocks}
The accessory parameters of the 5-point block are described by two monodromy equations read off from \eqref{moneq} at $n=5$ \cite{Alkalaev:2015lca}
\begin{equation}
\ba{c}
\label{mon_eq}
\left(I^{(5|1)}_{++}\right)^2+ I^{(5|1)}_{+-}I^{(5|1)}_{-+}  + 4\pi^{2} \tepsilon_{1}^{2} = 0\;,
\qquad
I^{(5|2)}_{+-}I^{(5|2)}_{-+} + 4\pi^{2} \tepsilon_{2}^{2} = 0 \;.
\ea
\end{equation}
These are two second order equations in $c_2, c_3$ that can be reduced to a fourth-order equation for one of these variables. There are four roots and their explicit form looks very massive at arbitrary dimensions $\epsilon_{1,2,3}, \tilde\epsilon_{1,2}$. And for that reason we simplify our  analysis by considering the case of identity blocks when one of dimensions $\tilde\epsilon_{1},\tilde\epsilon_{2}$ is set to zero. Then, the solutions can be given in a concise form. In the 5-point case there are two independent identity blocks, $\tilde{\epsilon}_{1}=0$ or $\tepsilon_{2} = 0$. Note that $\tilde{\epsilon}_{1}$ and $\tepsilon_{2}$ cannot be set to zero simultaneously because the fusion rules would imply that  $\epsilon_{3}=0$.

\paragraph{First identity block $\tilde{\epsilon}_1 =0$.} From the fusion rules \eqref{triangle} it follows that
\be
\label{con_mv2}
\tilde{\epsilon}_1 =0\;: \qquad \tepsilon_{2} = \epsilon_3\;, \qquad \epsilon_1 = \epsilon_2 \;,
\ee
and the monodromy equations \eqref{mon_eq} take the form
\begin{equation}
\ba{c}
\label{mon_eq_v2}
\left(I^{(5|1)}_{++}\right)^2 + I^{(5|1)}_{+-}I^{(5|1)}_{-+} = 0 \;,
\qquad
I^{(5|2)}_{+-}I^{(5|2)}_{-+} + 4\pi^{2} \epsilon^2_3 = 0 \;.
\ea
\end{equation}
Relevant roots of \eqref{mon_eq_v2} are given by
\be
\ba{c}
\dps c_1 =  \epsilon_1 P_1^{-1/\alpha} \left( 1-\alpha + \frac{2 \alpha P_1}{P_1 - P_2}\right)\;,
\quad
\dps c_2 = \epsilon_1 P_2^{-1/\alpha}\left(1-\alpha - \frac{2 \alpha P_1}{P_1 - P_2}\right)\;,
\quad
\dps c_3 = \epsilon_3 P_3^{-1/\alpha}\;,
\ea
\ee
where the expression for $c_1$ was obtained from \eqref{ccc}. Integrating  equations \eqref{acs1} we obtain the identity block
\be
\label{ccb_v2}
\ba{c}
f^{(1)}_{5}(z_{1,2,3}|\epsilon_{1,3}) = \epsilon_1(\alpha -1) \log P_1^{1/\alpha}  + \epsilon_1( \alpha - 1 ) \log P_2^{1/\alpha} - \epsilon_3 \log P_3^{1/\alpha} -
\vspace{-3mm}\\
\\
- 2 \epsilon_1 \log[P_1- P_2]\;.
\ea
\ee
The 5-point identity block \eqref{ccb_v2} can be represented as
\be
\label{vac5}
f^{(1)}_{5}(z_{1,2,3}|\epsilon_{1,3}) = f^{(1)}_{4}(z_{1,2}|\epsilon_{1}) +f_3(z_3|\epsilon_3)\;,
\ee
where the 4-point identity block and 3-point block are given by \eqref{4vac_s} and \eqref{3pt}, respectively.

The correspondence formula \eqref{fS} takes the form
\be
f^{(1)}_{5}(z_{1,2,3}|\epsilon_{1,3}) = - L^{(2)}_{\wdisk}(w_1,w_2|\epsilon_{1}) + i\epsilon_1 w_1 + i\epsilon_1 w_2 + i \epsilon_3 w_3\;,
\ee
where $L^{(2)}_{\wdisk}(w_1,w_2|\epsilon_1)$ is given by \eqref{arc}. Let us expand on this formula. A holographic Steiner tree dual to $5$-point block is $N=4$ graph (b) on Fig. \bref{N4}. Cutting the inner edge ($\tepsilon_1 =0$) yields the disconnected graph (a) on Fig. \bref{N2}. On the CFT side this decomposition is reflected in \eqref{vac5}.  Then, the total length of the holographic $N=4$ Steiner tree $L^{(4)}_{\wdisk} \to L^{(2)}_{\wdisk}$ because the contribution from the radial edge is zero, \eqref{radial}.

\paragraph{Second identity block $\tepsilon_{2}=0$.} The fusion rules \eqref{triangle} constrain the dimensions as follows
\be
\label{con_mv1}
\tepsilon_2 =0\;: \qquad \tilde{\epsilon}_{1} = \epsilon_3  \;,
\ee
and the monodromy equations \eqref{mon_eq} take the form
\begin{equation}
\ba{c}
\label{mon_eq_v1}
\left(I^{(5|1)}_{++}\right)^2 + I^{(5|1)}_{+-}I^{(5|1)}_{-+} + 4\pi^{2} \epsilon^2_3 = 0 \;,
\qquad
I^{(5|2)}_{+-}I^{(5|2)}_{-+} = 0 \;.
\ea
\end{equation}
The equations can be explicitly solved reduced as
\be
\label{acc_v}
\ba{c}
\dps c_1 = P_1^{-1/\alpha}\left(\epsilon_1(1-\alpha) + \frac{\alpha P_1 (\epsilon_1 + \epsilon_2 - \epsilon_3)}{P_1 - P_2} + \frac{\alpha P_1 (\epsilon_1 - \epsilon_2 + \epsilon_3)}{P_1 - P_3} \right)\;,
\vspace{-2mm}
\\
\\
\dps c_2 = P_2^{-1/\alpha}\left(\epsilon_2(1-\alpha) - \frac{\alpha P_2 (\epsilon_1 + \epsilon_2 - \epsilon_3)}{P_1 - P_2} - \frac{\alpha P_2 (\epsilon_2 + \epsilon_3 - \epsilon_1)}{P_3 - P_2} \right)\;,
\vspace{-2mm}
\\
\\
\dps c_3 = P_3^{-1/\alpha}\left(\epsilon_3(1-\alpha) - \frac{\alpha P_3 (\epsilon_1 - \epsilon_2 + \epsilon_3)}{P_1 - P_3} - \frac{\alpha P_3 (\epsilon_2 + \epsilon_3 - \epsilon_1)}{P_2 - P_3} \right)\;.
\ea
\ee
Then, the respective 5-point identity block is given by
\be
\label{ccb_v}
\ba{c}
f^{(2)}_{5}(z_{1,2,3}|\epsilon_{1,2,3} ) = \epsilon_1( \alpha - 1 ) \log P_1^{1/\alpha} + \epsilon_2( \alpha - 1 ) \log P_2^{1/\alpha} + \epsilon_3( \alpha - 1 ) \log P_3^{1/\alpha} -
\\
\\
-(\epsilon_1 + \epsilon_2 - \epsilon_3)\log(P_1 - P_2) - (\epsilon_1 - \epsilon_2 + \epsilon_3) \log(P_1 - P_3) - (\epsilon_3 + \epsilon_2 -\epsilon_1) \log(P_2 - P_3)\;.
\ea
\ee
Using the correspondence formula \eqref{fS} we can explicitly see that
\be
\label{bulk5pt}
f^{(2)}_{5}(z_{1,2,3}|\epsilon_{1,2,3} ) = - L^{(3)}_{\wdisk}(w_1,w_2,w_3)  + i \epsilon_1 w_1+ i \epsilon_2 w_2 + i \epsilon_3 w_3\;,
\ee
where $L^{(3)}_{\wdisk}(w_1,w_2,w_3)$ is given by the $w$-independent part of the length function of the holographic $N=3$ Steiner tree \eqref{vacuuma_5pt_length}, see (c) on Fig. \bref{N3pic}.

\subsection{6-point identity  blocks}

Let us consider the 6-point identity block obtained by setting $\tilde\epsilon_3=0$. (Other cases $\tepsilon_1 =0$ or $\tepsilon_2=0$ are dual to disconnected Steiner trees and will be discussed below.) From the fusion rules \eqref{triangle} it follows that
\be
\tilde{\epsilon}_3 =0\;: \qquad  \tilde{\epsilon}_2 = \epsilon_4\;.
\ee
In this case, the monodromy equations \eqref{moneq} take the form
\begin{equation}
\label{mon_eq6pt}
\ba{c}
\left(I^{(6|2)}_{++}\right)^2 + I^{(6|1)}_{+-}I^{(6|1)}_{-+} + 4\pi^{2} \tilde{\epsilon}^2_1 = 0 \;, \qquad \left(I^{(6|1)}_{++}\right)^2 + I^{(6|2)}_{+-}I^{(6|2)}_{-+} + 4\pi^{2} \epsilon^2_4 = 0 \;,
\\
\\
I^{(6|3)}_{+-}I^{(6|3)}_{-+} = 0 \;.
\ea
\end{equation}
The second equation of the system follows from the third one, so it can be reduced to the system of two equations. One of these equations is linear, the second one is quadratic, and, therefore,  the system can be solved exactly. These equations fix any two accessory parameters of the  three original ones. For example, these are $c_2$ and $c_3$, while $c_1$ can be obtained from \eqref{ccc}.

To simplify our calculations we consider the case $\epsilon_1 = \epsilon_2$, $\epsilon_3 = \epsilon_4$. Then, the monodromy system \eqref{mon_eq6pt} is solved as
$$
\dps c_1 =   P_1^{-1/\alpha}\left[\epsilon_1 + \alpha\left(\frac{\epsilon_1 (P_1 + P_2)}{P_1 -P_2} - \frac{\tilde\epsilon_1 P_1\sqrt{(P_1 - P_4)(P_2 - P_4)(P_1 - P_3)(P_2 - P_3)}}{(P_1 - P_2)(P_1 - P_3)(P_1 - P_4)}\right)\right]\;,
$$
$$
c_2 =  P_2^{-1/\alpha}\left[\epsilon_1 - \alpha\left(\frac{\epsilon_1 (P_1 + P_2)}{P_1 -P_2} - \frac{\tilde\epsilon_1 P_2\sqrt{(P_1 - P_4)(P_2 - P_4)(P_1 - P_3)(P_2 - P_3)}}{(P_1 - P_2)(P_2 - P_3)(P_2 - P_4)}\right)\right]\;,
$$
$$
c_3 = P_3^{-1/\alpha}\left[\epsilon_3 + \alpha\left(\frac{\epsilon_3 (P_3 + P_4)}{P_3 -P_4} - \frac{\tilde\epsilon_1 P_3\sqrt{(P_1 - P_4)(P_2 - P_4)(P_1 - P_3)(P_2 - P_3)}}{(P_3 - P_4)(P_2 - P_3)(P_1 - P_3)}\right)\right]\;,
$$
\be
c_4 = P_4^{-1/\alpha}\left[\epsilon_3  - \alpha\left( \frac{\epsilon_3 (P_3 + P_4)}{P_3 -P_4} - \frac{\tilde\epsilon_1 P_4 \sqrt{(P_1 - P_4)(P_2 - P_4)(P_1 - P_3)(P_2 - P_3)}}{(P_3 - P_4)(P_2 - P_4)(P_1 - P_4)}\right)\right]\;.
\ee
Integrating \eqref{acs1} we obtain the 6-point identity block
\be
\ba{c}
\dps f^{(3)}_{6} (z_{1,2,3,4}|\epsilon_{1,3},\tilde\epsilon_1) = \epsilon_1( \alpha - 1 ) (\log P_1^{1/\alpha} + \log P_2^{1/\alpha})+ \epsilon_3( \alpha - 1 ) (\log P_3^{1/\alpha} + \log P_4^{1/\alpha}) -
\\
\\
- 2\epsilon_1 \log(P_1- P_2) - 2\epsilon_3\log(P_3 - P_4) +
\\
\\
\dps + \tilde{\epsilon}_1 \log\left[ \frac{2(P_1 - P_4)(P_2 - P_3)}{(P_1 - P_2)(P_3 - P_4)}  -\frac{2\sqrt{(P_1 - P_3)(P_2 - P_3)(P_1 - P_4)(P_2 - P_4)}}{(P_1 - P_2)(P_3 - P_4)} + 1\right]\;.
\ea
\ee
This expression satisfies the  correspondence formula \eqref{fS}
\be
f^{(3)}_{6} (z_{1,2,3,4}|\epsilon_{1,3},\tilde\epsilon_1)  = -L^{(4)}_{\wdisk}(w_{1,2,3,4}|\epsilon_{0,1,3}) + i \epsilon_1 w_1+ i \epsilon_1 w_2 + i \epsilon_3 w_3 + i \epsilon_3 w_4 \;,
\ee
where  $L^{(4)}_{\wdisk}(w_1,w_2,w_3,w_4)$ is the $w$-independent part of the length of the $N=4$ ideal holographic Steiner tree \eqref{answ_br}, see (a) on Fig. \bref{N4}.

\section{Identity blocks and factorization}
\label{sec:factor}

Recall that an identity conformal block is obtained by choosing one of exchange channels to be an identity  operator so that its dimension is zero.\footnote{\label{footnote:vac} When all possible exchange channels are unit operators we  call such a block {\it vacuum}. Vacuum blocks are operational in calculating the entanglement entropy at large central charge \cite{Hartman:2013mia}.} From Section \bref{sec:cut} it follows that unifying a number of connected holographic Steiner trees we obtain a Steiner tree of the same holographic type but disconnected. However, this factorization property is far from evident when rephrased in CFT terms. Indeed, both the original (quantum)  and classical blocks with an identity exchange  channel do not in general factorize into two blocks. In this section we explicitly show that in the zeroth exchange limit perturbative classical blocks do factorize into a sum of other perturbative classical blocks.

\subsection{Factorization relation}
\label{sec:f_rel}

\begin{figure}[H]
\centering
\includegraphics[width=120mm]{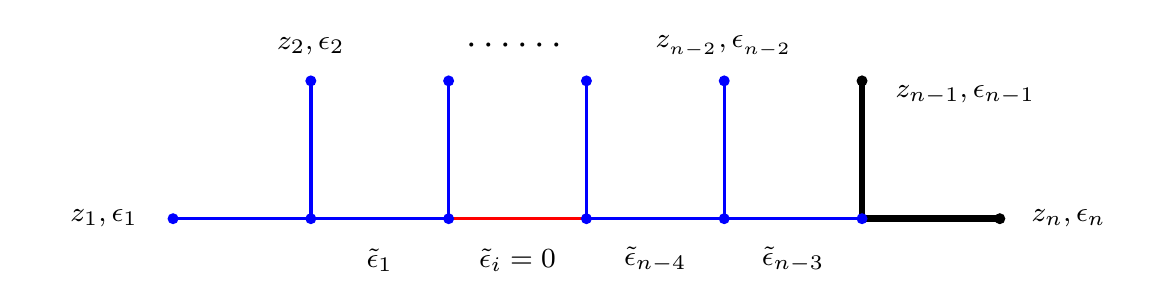}
\caption{$i$-th identity block: $\tilde\epsilon_i =0$, the identity channel is shown in red.  }
\label{blockID}
\end{figure}

Let $f_n(z|\alpha,\epsilon, \tepsilon)$ denote a perturbative $n$-point classical block and $f^{(i)}_n(z|\alpha,\epsilon, \tepsilon)$ denote a block obtained from $f_n(z|\alpha,\epsilon, \tepsilon)$ by setting $i$-th intermediate dimension to zero, i.e.,
\be
\label{iden}
f^{(i)}_n(z|\alpha,\epsilon, \tepsilon) \equiv  f_n(z|\alpha,\epsilon, \tepsilon)\big|_{\tepsilon_i = 0}\;\;.
\ee
We call such a function $f^{(i)}_n(z|\alpha,\epsilon, \tepsilon)$ an $i$-th {\it identity} $n$-point perturbative classical block, or, $i$-th identity block, for short. In this notation, $ f_n(z|\alpha,\epsilon, \tepsilon)\equiv f^{(0)}_n(z|\alpha,\epsilon, \tepsilon)$. Pictorially, the  identity block diagram is shown on Fig. \bref{blockID}. The interesting case is when the rightmost channel is an identity, $\tepsilon_{n-3}=0$. The respective diagram looks like that of the original $n$-point  classical block. Therefore, we call such an $n$-point identity  block  {\it maximal}.

We will prove the following factorization relation for the perturbative classical blocks
\be
\label{factor}
f^{(s)}_n(z|\alpha,\epsilon, \tepsilon) = f^{(s)}_{s+3}(z^\prime|\alpha,\epsilon^\prime, \tepsilon^\prime)+f^{(0)}_{n-s-1}(z^{\prime\prime}|\alpha,\epsilon^{\prime\prime}, \tepsilon^{\prime\prime})\;,
\ee
at $s = 1,2,...\,,n-3$, where the first term on the right-hand side is $(s+3)$-point maximal identity  block, and the second term is  $(n-s-1)$-point block. The $z$-dependence splits into two subsets as $z = (z^\prime,z^{\prime\prime})$, where $z^\prime = \{z_1,..., z_{s+1}\}$ and $z^{\prime\prime} = \{z_{s+2}, ..., z_{n-2}\}$. In the sequel, using the projective conformal invariance we always set $z_1 =0$. Note that the second subset of points starts with $z_{s+2}\neq 0$ that explains why we rederived the monodromy equation keeping the first point arbitrary. The classical dimensions in \eqref{factor} are also split into two subsets. Here, we have to take into account the fusion rules \eqref{triangle} that yield the following constraints
\be
\label{fusion_gen}
\tepsilon_s =0\;:\qquad\quad \tepsilon_{s-1} = \epsilon_{s+1}\;, \qquad \tepsilon_{s+1} = \epsilon_{s+2}\;.
\ee
Therefore, having $\epsilon = (\epsilon_1,...\,,\epsilon_{n-2})$ and $\tepsilon = (\tepsilon_1,...\,,\tepsilon_{n-3})$ on the left-hand side of \eqref{factor} we obtain on the right-hand side
\be
\ba{c}
\epsilon^{\prime} = (\epsilon_1,...\,,\epsilon_{s+1})\;, \qquad \epsilon'' = (\epsilon_{s+2},...\,,\epsilon_{n-2})\;,
\vspace{-2mm}
\\
\\
\tepsilon'=(\tepsilon_1...\,,\tepsilon_{s-2})\;, \qquad \tepsilon''=(\tepsilon_{s+2},...\,,\tepsilon_{n-3})\;,
\ea
\ee
with exception of $\epsilon^{\prime} = (\epsilon_1\,,\epsilon_1)$ in the case of $s=1$ identity block. A few comments are in order. In the bulk, maximal identity blocks correspond to ideal holographic Steiner trees (see Section \bref{sec:cut}). Also, we note that the factorization relation is non-trivial only for $s\leq n-4$. When $s=n-3$ the relation \eqref{factor} becomes the identity because $f^{(0)}_{2}(z|\alpha,\epsilon, \tepsilon) = 0$, see the end of Section \bref{sec:HL}.
\begin{figure}[H]
\centering
\includegraphics[width=100mm]{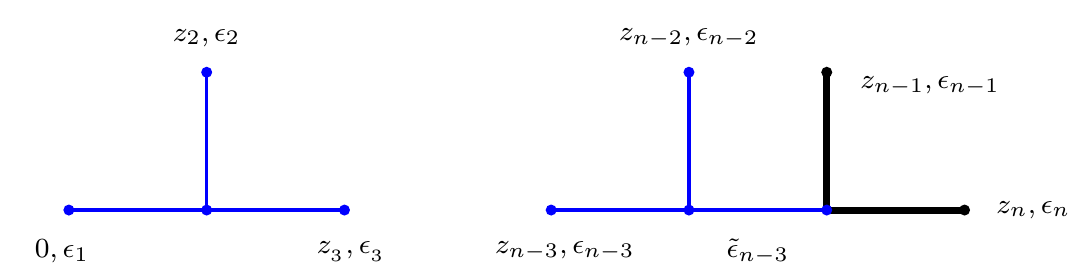}
\caption{The factorization relation for the identity block from Fig. \bref{blockID}.}
\label{fig:factor}
\end{figure}

A given $n$-point identity block \eqref{iden} is a particular value of the original block and, therefore, satisfies the $n$-point monodromy system \eqref{moneq}. Here, to simplify our presentation we identify knowing the block function with knowing its accessory parameters. A-priori, it is not evident whether a given identity block can be represented as a sum of two other blocks. For this to happen it is necessary that accessory parameters corresponding to those two blocks satisfy the original $n$-point monodromy system \eqref{moneq} where the dimensions are subjected to the fusion constraints \eqref{fusion_gen}. On the other hand, the accessory parameters of the two blocks satisfy their own monodromy systems with less number of equations. Meanwhile, counting  accessory parameters on both sides of the factorization relation \eqref{factor} we find out that their numbers are different.  Therefore, the problem is to show that equating one of intermediate dimensions to zero the original monodromy system indeed has a solution corresponding to two independent subsystems indicated in the factorization relation.

To prove the factorization relation \eqref{factor} we first  consider the case $k=1$. We explicitly demonstrate our technique and then generalize the proof to  the case of arbitrary $k$ in  Appendix \bref{app:factor}. At $s=1$ in \eqref{factor} the factorization relation reads  $f^{(1)}_n(z|\epsilon, \tepsilon) = f^{(1)}_{4}(z^\prime|\alpha,\epsilon^\prime, \tepsilon^\prime)+f^{(0)}_{n-2}(z^{\prime\prime}|\alpha,\epsilon^{\prime\prime}, \tepsilon^{ \prime\prime})$ that must follow from the general monodromy  equations \eqref{moneq} at $\tepsilon_1 = 0$. The fusion constraints \eqref{fusion_gen} in this case are
\be
\label{fusion4pt}
\tepsilon_1 = 0\;:\qquad\quad \epsilon_1 = \epsilon_2\;, \qquad \tepsilon_2 = \epsilon_3\;.
\ee

Let us consider the system \eqref{moneq} for the 4-point block $f_4(0,z_2|\alpha,\epsilon_2,\epsilon_2, 0)$ which is the identity block, see Section \bref{sec:4pt}. In this case, there is a single  equation
\be
\label{moneq4}
I_{+-}^{(4|1)}\,I_{-+}^{(4|1)}+\left(I_{++}^{(4|1)}\right)^2 =0\;,
\ee
where from \eqref{Is} we find that $I_{++}^{(4|1)} = 0$, and
\be
\ba{l}
\dps I_{+-}^{(4|1)} =\frac{2\pi i}{\alpha}\Big[ \alpha \epsilon_2 +c_2(1-z_2) - \epsilon_2 - (1-z_2)^\alpha(c_2(1-z_2)-\epsilon_2(1+\alpha))\Big]\;,
\vspace{-2mm}
\\
\\
\dps I_{-+}^{(4|1)} =-\frac{2\pi i}{\alpha}\Big[ -\alpha \epsilon_2 +c_2(1-z_2) - \epsilon_2 - (1-z_2)^{-\alpha}(c_2(1-z_2)-\epsilon_2(1-\alpha))\Big]\;.
\ea
\ee
Noting that $I_{-+}^{(4|1)} = (1-z_2)^{-\alpha}I_{+-}^{(4|1)}$ we find  out that the equation \eqref{moneq4} factorizes, and, therefore, is equivalent to
\be
\label{id4}
I_{+-}^{(4|1)} = 0\;.
\ee
In particular, this linear equation in $c_2$ can be directly solved to yield the 4-point identity block function.

Now, we analyze the monodromy  system \eqref{moneq} where the first point is set $z_1=0$, and the dimensions are subjected to the fusion conditions \eqref{fusion4pt}. There are $n-3$ accessory parameters $c_2, ...\,, c_{n-2}$, and let the first  parameter $c_2$ satisfy the 4-point identity block system \eqref{id4}. We show that of $n-3$ equations in \eqref{moneq} the first two are identically satisfied while the remaining $n-5$ equations are non-trivial and describe $(n-2)$-point block.

\vspace{-2mm}

\paragraph{1st equation.} Consider the first equation $k=1$ in the $n$-point system \eqref{moneq}. Recalling the constraints \eqref{fusion4pt} we obtain
\be
\label{moneq4n}
I_{+-}^{(n|1)}\,I_{-+}^{(n|1)}+\left(I_{++}^{(n|1)}\right)^2 =0\;,
\ee
where
\be
\label{Isk1}
\ba{c}
\dps
I^{(n|1)}_{+-}  =\frac{2\pi i}{\alpha}\,\Big[\alpha \epsilon_2+\sum_{i=2}^{n-2}(c_i(1-z_i)-\epsilon_i)-(1-z_2)^\alpha(c_2(1-z_2)-\epsilon_2(1+\alpha))\Big],
\vspace{-3mm}

\\
\\
\dps
I^{(n|1)}_{-+}  = I^{(n|1)}_{+-}\big|_{\alpha \rightarrow -\alpha}\,,
\qquad\;\;
I^{(n|1)}_{++} = \frac{2\pi i}{\alpha}\,\sum_{i=3}^{n-2}\big[c_i (1-z_i)-\epsilon_i\big]\,.
\ea
\ee
By assumption, the parameter $c_2$ satisfies \eqref{id4} so that we can substitute that condition into \eqref{Isk1} and  find out that
\be
I^{(n|1)}_{+-}    \approx I^{(n|1)}_{++}\;,
\qquad
I^{(n|1)}_{-+}  \approx -I^{(n|1)}_{++}\;,
\ee
where the weak equality $\approx$ means that we used   \eqref{id4}. It immediately follows that the equation \eqref{moneq4n} is identically satisfied.

\vspace{-2mm}

\paragraph{2nd equation.} The $k=2$ equation in the $n$-point system \eqref{moneq} is given by
\be
\label{secondeq}
I_{+-}^{(n|2)}\,I_{-+}^{(n|2)}+\left(I_{++}^{(n|2)}\right)^2 + 4\pi^2 \epsilon_3^2 = 0\;,
\ee
where $I_{+-}^{(n|2)}$ has been evaluated using the condition \eqref{id4},
\be
\label{Isk2}
\ba{c}
\dps
I^{(n|2)}_{+-}  \approx \frac{2\pi i}{\alpha}\Big[\sum_{i=3}^{n-2}(c_i(1-z_i)-\epsilon_i)-(1-z_3)^\alpha(c_3(1-z_3)-\epsilon_3(1+\alpha))\Big],
\vspace{-2mm}
\\
\\
\dps
I^{(n|2)}_{-+}  = I^{(n|2)}_{+-}\big|_{\alpha \rightarrow -\alpha}\,,
\qquad\;\;
I^{(n|2)}_{++} = \frac{2\pi i}{\alpha}\,\sum_{i=4}^{n-2}\big[c_i (1-z_i)-\epsilon_i\big]\,.
\ea
\ee

Let us now consider the second factor $f^{(0)}_{n-2}(z^{\prime\prime}|\alpha,\epsilon^{\prime\prime}, \tepsilon^{ \prime\prime})$ in the factorization condition. It depends on points $z^{\prime\prime}=(z_3,... , z_{n-2})$, where the first point $z_3\neq 0$  and the associated accessory parameters are $c_3, ..., c_{n-2}$. It is known that the parameters of the $(n-2)$-point monodromy system are linearly dependent
\be
\label{sumacc}
\sum_{i=3}^{n-2}\big[c_i (1-z_i)-\epsilon_i\big] = 0\;,
\ee
that directly follow from \eqref{ccc} by relabelling indices. Using \eqref{sumacc} and denoting $x_3 = \epsilon_3(1-z_3)-\epsilon_3$ we can rewrite  \eqref{Isk2} in terms of $x_3$ so that the monodromy equation  \eqref{secondeq} reads
\be
\frac{4\pi^2}{\alpha^2}(x_3 - \alpha \epsilon_3)(x_3 + \alpha \epsilon_3) - \frac{4\pi^2}{\alpha^2}x_3^2 + 4\pi^2 \epsilon_3^2 = 0\;,
\ee
which is again identically satisfied.
\paragraph{Other equations.} The remaining equations in the $n$-point system \eqref{moneq} are
\be
\label{remeq}
I_{+-}^{(n|k)}\,I_{-+}^{(n|k)}+\left(I_{++}^{(n|k)}\right)^2  + 4\pi^2 \tilde\epsilon^2_{k}=0\;,
\qquad k = 3,...\,, n-3\,,
\ee
where using the condition \eqref{id4} and the relation between accessory parameters \eqref{sumacc} we get
\be
\label{Is3}
\ba{c}
\dps
I^{(n|k)}_{+-}  \approx\frac{2\pi i}{\alpha}\Big[-\sum_{i=3}^{k+1}(1-z_i)^\alpha(c_i(1-z_i)-\epsilon_i(1+\alpha))\Big],
\vspace{-3mm}

\\
\\
\dps
I^{(n|k)}_{-+}  = I^{(n|k)}_{+-}\big|_{\alpha \rightarrow -\alpha}\,,
\qquad\;\;
I^{(n|k)}_{++} = \frac{2\pi i}{\alpha}\sum_{i=k+2}^{n-2}\big[c_i (1-z_i)-\epsilon_i\big]\,.
\ea
\ee
In order for this system to describe the $(n-2)$-point block we have to show that $I^{(n|k)}_{+-}$ should take the form \eqref{Is} where $k\to k+2$. That would effectively mean that the monodromy system \eqref{moneq} describes $(n-2)$-point block with points enumerated as $3,4,...\;$.  At the same time, $I^{(n|k)}_{++} $ is already (modulo shifting $k$) of the required form. We observe that the $i=3$ term in $I^{(n|k)}_{+-}$ \eqref{Is3} can be represented as
$-(1-z_3)^\alpha (c_3(1-z_3) -\epsilon_3(1+\alpha)) = (1-z_3)^\alpha(\alpha\epsilon_3 +\sum_{i=4}^{n-2}(c_i (1-z_i)-\epsilon_i))$, where we used \eqref{sumacc}. Substituting this relation back into \eqref{Is3} we reproduce \eqref{Is}. We conclude that the equation system \eqref{remeq} indeed describes accessory parameters of the general $(n-2)$-point block.

Thus, we have shown that the factorization condition \eqref{factor} is satisfied when $s=1$. The proof can be straightforwardly extended to arbitrary $s$, see Appendix \bref{app:factor}. The general idea is to use the monodromy equations for  the maximal $(s+3)$-point identity block and then, along with linear relation for the accessory parameters of the general $(n-s-1)$-point block  and the fusion constraints, to show that the original $n$-point monodromy equations reduce to the $(n-s-1)$-point monodromy system thereby proving the factorization relation \eqref{factor}.

\subsection{Multiple identity blocks}

Two  blocks on the right-hand side of the factorization condition \eqref{factor}  can be further factorized by setting other intermediate dimensions to zero. Factorization of the second factor $f^{(0)}_{n-2}(z^{\prime\prime}|\alpha,\epsilon^{\prime\prime}, \tepsilon^{ \prime\prime})$ is obvious due to the same factorization relation.  It is more interesting to consider  how the first factor $f^{(s)}_{s+3}(z^\prime|\alpha,\epsilon^\prime, \tepsilon^\prime)$ (maximal identity block) factorizes.

Let us suppose now that one of intermediate dimensions of the maximal identity block is set to zero, i.e. $\tepsilon_m =0$  for some $m\in \{1,2,...,s-1\}$. The fusion rules, in this case, yield
$\tepsilon_{m-1} = \epsilon_{m+1}$, $\tepsilon_{m+1} = \epsilon_{m+2}$, cf. \eqref{fusion_gen}.

The factorization relation for the maximal identity block is given by
\be
\label{factor_max}
f^{(s)}_{s+3}(z|\alpha,\epsilon, \tepsilon) = f^{(m)}_{m+3}(z^\prime|\alpha,\epsilon^\prime, \tepsilon^\prime)+f^{(s-m-1)}_{s-m+2}(z^{\prime\prime}|\alpha,\epsilon^{\prime\prime}, \tepsilon^{\prime\prime})\;,
\ee
where on the right-hans side we have two maximal identity blocks of lower ranks, both the coordinates and dimensions are properly split. The proof goes along  the same lines as for the original  factorization relation  \eqref{factor}.

It is clear that equating intermediate dimensions to zero can  be continued to produce more identity blocks that can be denoted as
\be
f^{(a,b,...,c)}_n(z|\alpha,\epsilon, \tepsilon) \equiv f_n(z|\alpha,\epsilon, \tepsilon)\big|_{\tepsilon_a,\tepsilon_b,...,\tepsilon_c  = 0}\;\;,
\ee
where  integers $a,b,...,c = 1,..., n-3$ label identity exchange channels. This process, however, is terminated at some stage  because there is a finite number of exchange channels, and, moreover, the fusion rules forbid equating intermediate dimensions to zero simultaneously.\footnote{If $\tepsilon_{i} = \tepsilon_{i+1} = 0$ then $\epsilon_{i+2}=0$ meaning that the original $n$-point block is reduced to the $(n-1)$-point block. This corresponds to the cut rule \eqref{cut1} of Section \bref{sec:cut}. The factorization we discuss here keeps the number of points $n$ intact. } The extreme case is when maximum  possible number of intermediate dimensions is set to zero: (a) $\tepsilon_{2i+1} = 0$ for $i=0,1,2,...$, (b) $\tepsilon_{2i} = 0$ for $i=1,2,...$ (see our comments in the footnote \bref{footnote:vac}). Then, the original block is factorized into a sequence of 4-point identity blocks, and, possibly,  3-point block along with one of 5-point identity  blocks.\footnote{Similar but  different factorization in other  OPE channels were discussed  in \cite{Banerjee:2016qca,Hirai:2018jwy}.}

\begin{figure}[H]
\centering
\includegraphics[width=125mm]{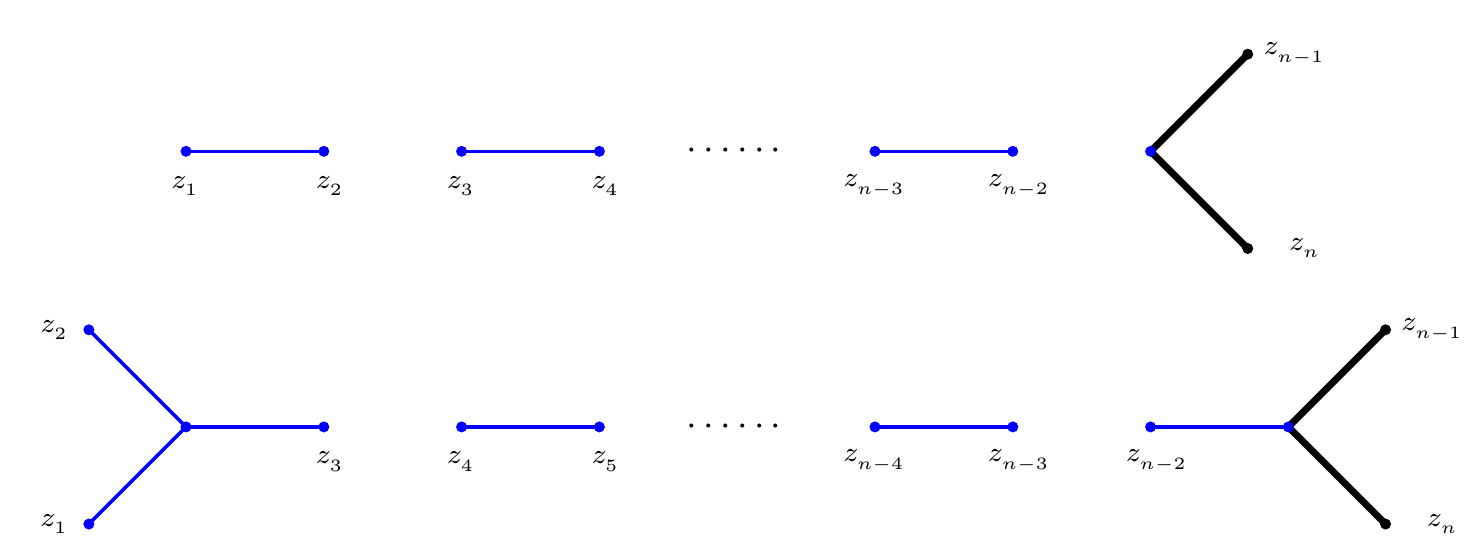}
\caption{Vacuum blocks with even $n$. The upper diagram corresponds to the disconnected Steiner tree (a) on Fig. \bref{dis}.}
\label{even}
\end{figure}

Let $n$ be even. Then, using the fusion rules \eqref{fusion_gen} we find two decompositions for identity blocks according to the subsets  (a) and (b) above,
\be
\label{id_factor1}
f^{(1,3,...,n-3)}_n(z|\alpha,\epsilon, \tepsilon) = \sum_{i=1}^{\frac{n-2}{2}} f^{(1)}_{4}(z_{2i-1}, z_{2i}|\alpha,\epsilon_{2i-1})\;,
\ee
\be
\dps f^{(2,4,...,n-4)}_n(z|\alpha,\epsilon, \tepsilon) = f_5^{(2)}(z_{1,2,3}|\alpha, \epsilon_{1,2,3})+f_3(z_{n-2}|\alpha, \epsilon_{n-2})+\sum_{i=2}^{\frac{n-4}{2}} f^{(1)}_{4}(z_{2i}, z_{2i+1}|\alpha,\epsilon_{2i})\;,
\ee
where 4-point identity blocks are given by \eqref{4vac_s}, 3-point block is given by \eqref{3pt}, and 5-point identity block is given by \eqref{ccb_v}, see Fig. \bref{even}.

\begin{figure}[H]
\centering
\includegraphics[width=135mm]{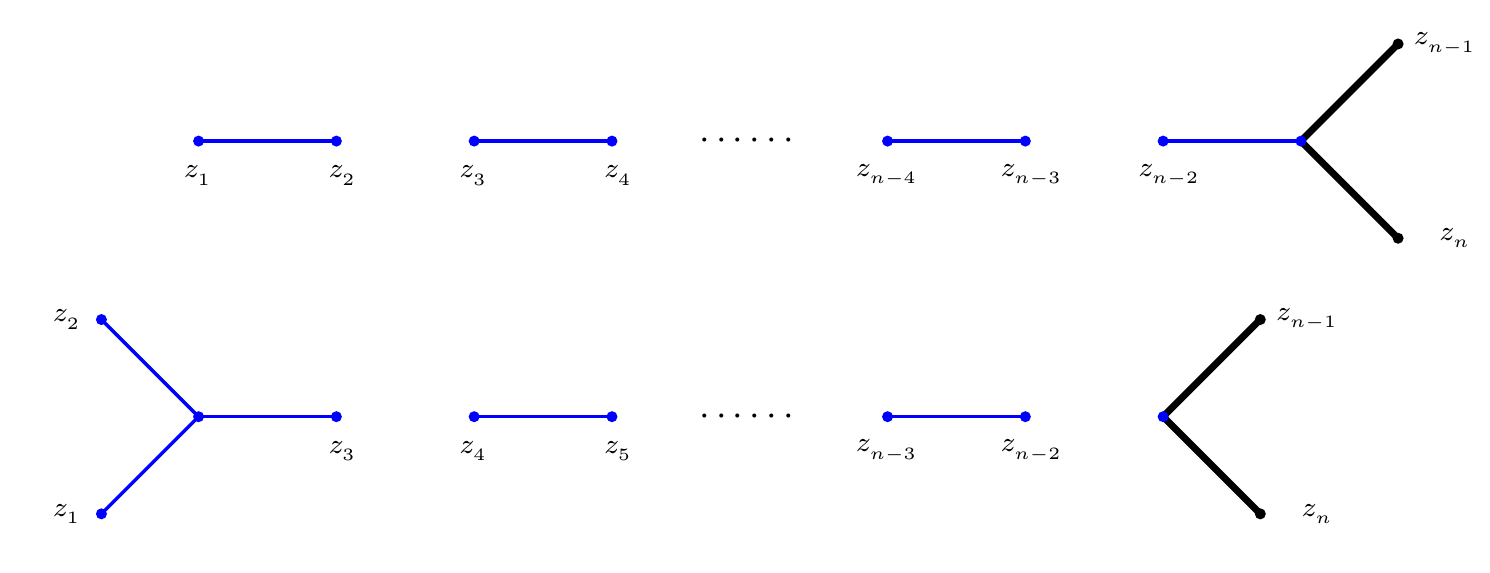}
\caption{Vacuum blocks with odd $n$. The bottom diagram corresponds to the disconnected Steiner tree (b) on Fig. \bref{dis}.}
\label{odd}
\end{figure}

The analogous decompositions of identity blocks hold in the odd $n$ case,
\be
\label{odd1}
f^{(1,3,...,n-4)}_n(z|\alpha,\epsilon, \tepsilon) =f_3(z_{n-2}|\alpha, \epsilon_{n-2})+ \sum_{i=1}^{\frac{n-3}{2}} f^{(1)}_{4}(z_{2i-1}, z_{2i}|\alpha,\epsilon_{2i-1})\;,
\ee
\be
f^{(2,4,...,n-3)}_n(z|\alpha,\epsilon, \tepsilon) =  f^{(2)}_{5}(z_{1,2,3}|\alpha,\epsilon_{1,2,3})+\sum_{i=2}^{\frac{n-3}{2}} f^{(1)}_{4}(z_{2i}, z_{2i+1}|\alpha,\epsilon_{2i})\;,
\ee
see Fig. \bref{odd}. The expression for the 5-point identity block \eqref{vac5} is a particular example of decomposition  \eqref{odd1}.

\vspace{-6mm}

\vspace{-6mm}

To conclude this section let us note that the logic of the block/length correspondence can be inverted in the sense that having established  the factorization relation for perturbative classical blocks on the  CFT side we immediately conclude that in the bulk  the graph theory realization should give rise to graphs of particular type where cutting one inner edge yields a disconnected graph. On the other hand, since the full classical block (already the second order  correction) has no such a factorization property it means that the dual graphs have loops. It is natural, because exchange channels in the leading heavy-light approximation are given by one (primary) state and considering sub-leading terms  yields more (secondary) states, and, therefore, more complicated dual graphs.

\section{Conclusion}

In this paper we reformulated the heavy-light regime of the semiclassical $\dual$ correspondence as the correspondence between weighted Steiner trees in the hyperbolic geometry and perturbative classical blocks. The novel part here is a geometric view on the geodesic networks in the bulk space. This approach allows us to find the total lengths more effectively  because we use the standard $(z, \bar z)$ parameterization of the hyperbolic spaces and interaction vertices of the particle's worldlines are described as the generalized Fermat-Torricelli points. In the general case of $n$-point blocks these two observations help to explicitly integrate a part of complicated algebraic equations of motion arising within the worldline formulation \cite{Alkalaev:2015wia}. To demonstrate our technique we have explicitly found lengths of particular Steiner trees dual to different $n$-point perturbative blocks with $n=3,4,5,6$.

Formalizing the bulk description of conformal blocks in the heavy-light approximation as holographic Steiner trees helps to establish the factorization property  of the conformal blocks. Indeed, having the special graph theory in the bulk space we can discuss its various geometric properties including  cuts and connectivity. In this way, we immediately obtain a simple description of (dis)connected Steiner trees. On the CFT side,  connectivity and cuts are directly translated into  factorization  and identity blocks. However, the factorization property is far from evident because  block functions are obtained by integrating the accessory parameters which in their turn satisfy complicated monodromy equations. Nonetheless, we were able to prove the factorization relations purely in CFT terms without referring to the bulk formulation. As a by-product,  we classify all identity blocks associated to a given $n$-point perturbative classical block and discuss their realization in terms of the Steiner trees. By way of illustration, we have explicitly found the accessory parameters and the corresponding block functions in the $n=3,4,5,6$ cases that supplement our analysis of the holographic  Steiner trees in the bulk.

As a concluding comment let us note that from the bulk/boundary perspective the perturbative classical blocks can be thought of as emergent objects. Indeed, one can view conformal blocks as physical quantities  because these form a basis in the space of $n$-point correlation functions. On the other hand, we see that it is the factorization properties  of the conformal blocks that ensure  their identification with purely abstract  objects like trees and forests in the graph theory.

It would be interesting to extend our analysis of the Steiner trees in the hyperbolic geometry in several directions. For example, one can consider the inverse Steiner problem: given FT points and endpoints, to characterize the weights  which minimize the total length. For $N=3$ graph the answer is positive \cite{10.2307/2695644}. For $N\geq 4$ Steiner trees the inverse problem has not been analyzed. Moreover, we hope that having defined holographic Steiner trees as those inscribed into $N$-gons with $N-1$ ideal vertices we might use many relevant results from two-dimensional hyperbolic geometry and, thus, to make progress in calculating total lengths of general trees.   Also, it would be interesting to consider other OPE channels in this context, see e.g. \cite{Banerjee:2016qca,Hulik:2016ifr}. Hopefully, explicit expressions for the total length of the Steiner trees like (c) on Fig. \bref{N3pic} or (a) on Fig. \bref{N4} could be interesting in the context of studying the entanglement entropy phenomena,  see e.g.  \cite{Banerjee:2016qca,Hirai:2018jwy,Ruggiero:2018hyl,Kusuki:2018wpa}.


\noindent \textbf{Acknowledgements.} M.P. would like to thank Kirill Stupakov for discussions. The  work was supported by RFBR grant No 18-02-01024.

\appendix

\section{Boundary regularization}
\label{app:A}

We can approach the boundary in different ways. The most natural is when a given point flows along the geodesic intersecting the boundary somewhere. This is the shorts path and the regulator can be introduced as the inverse geodesic length. However, we use a different regularization, when the angle coordinate of a given point remain fixed, while the radius tends to $1$ as $r \sim e^{-\varepsilon}$, where $\varepsilon$ is the boundary cut-off parameter. We use this prescription because the boundary attachments of outer edges of holographic Steiner trees  depend only on angles which are convenient to be kept fixed.

An edge between two boundary points $z_1=\exp[iw_1]$ and $z_2=\exp[iw_2]$ has an infinite length as most evident from  the general formula \eqref{length} represented as
\begin{equation}
\label{eq3}
L_{\mathbb{D}}(w_1, w_2)= \log \frac{1+u}{1-u}\;, \qquad \text{where}\qquad u = 1\;.
\end{equation}
In order to regularize the logarithm we introduce the boundary cut-off  as  $z_1 = \exp[-\varepsilon + i w_1]$ and $z_2=\exp[-\varepsilon+i w_2]$ at $\varepsilon \to +0$. Let $w_2 - w_1 \geq 0$. Then, defining
$u=A/B$,   where
\be
\label{AB}
A=2\exp[-\varepsilon]\sin\frac{w_2 - w_1}{2} \;,
\quad
B = \sqrt{(1- \exp[-2\varepsilon]\cos (w_2 - w_1))^2+\exp[-4\varepsilon]\sin^2 (w_2 - w_1)}\;,
\end{equation}
we find that the length function can be represented as
\begin{equation}
\label{eq10}
L_{\mathbb{D}}(w_1, w_2)=\log\frac{(B+A)^2}{B^2-A^2} \;.
\end{equation}
It is remarkable that in this form the denominator depends only on $\varepsilon$ that allows us to isolate the divergence. Indeed, using \eqref{AB} we find
\be
\label{eq11}
\ba{c}
\dps(B+A)^2 = 16 \sin^2 \frac{w_2 - w_1}{2} -32\sin^2 \frac{w_2 - w_1}{2} \varepsilon+ \cO(\varepsilon^2)\;,
\vspace{-3mm}
\\
\\
B^2-A^2= \left(1 - \exp[-2 \epsilon]\right)^2 = 4\varepsilon^2 + \cO(\varepsilon^2)\;.
\ea
\ee
The length function represented as a finite part plus logarithmic divergence is given by
\begin{equation}\label{eq13}
L_{\mathbb{D}}(w_1, w_2)= \log\left[4\sin^2\frac{w_2 - w_1}{2}\right] - 2\log \varepsilon + \cO(\epsilon)\;.
\end{equation}
Thus, we find that a regularized length is defined to be the leading term in the decomposition \eqref{eq13}.

Now, consider an edge between a boundary point $z_{1}=\exp[ - \varepsilon + i w]$ where $\varepsilon \to +0$  and a point inside the disk, $z_{2}=r \exp[i\varphi]$. Similarly to the previous case we define $u = A/B$, where
\begin{equation}
\label{appA_2}
\ba{l}
\dps
A = \sqrt{r^2 + \exp[ - 2\varepsilon] -2r\exp[ - \varepsilon]\cos(w-\varphi)}\;,
\quad
B =\sqrt{ 1+r^2 \exp[-2\varepsilon] - 2r \exp[-\varepsilon]\cos(w-\varphi)}\;.
\ea
\end{equation}
Representing the length function as in  \eqref{eq10} we find
\be
\ba{c}
(B+A)^2 = 4 \left(r^2-2 r \cos(\varphi-w)+1\right) + \cO(\varepsilon)\;,
\vspace{-3mm}
\\
\\
B^2 - A^2 =   2\left(1-r^2\right)\varepsilon-2\left(1- r^2\right) \varepsilon^2 + \cO(\varepsilon^3)\;.
\ea
\ee
Isolating the logarithmic divergence we obtain
\be
\label{regL}
L_{\mathbb{D}}(w, r, \varphi) = \log \frac{ \left(r^2-2 r \cos(\varphi-w)+1\right)}{1-r^2} + \log 2 - \log \varepsilon + \cO(\varepsilon)\;.
\ee

\section{Details of calculations}
\label{app:N34}

\paragraph{N=3 trees in Section \bref{sec:N3}.}  In what follows we calculate the length of the Steiner tree (a)  on Fig. \bref{N3pic}. The boundary points $z_{1,2}$, the point inside the disk $z_{0}$,  and the FT point $z_{_{FT}}$ are given in polar coordinates,
\begin{equation}
\label{eq2_1}
z_{1}=\exp[iw_1-\varepsilon]\;,
\quad
z_{2}=\exp[iw_2-\varepsilon]\;,
\quad z_{0} = r_0\exp[i w_0] \;,
\quad z_{_{FT}}=r\exp[i \varphi] \;.
\end{equation}
From the general circle equation \eqref{circle} we find that the edge connecting points $z_{0}$ and $ z_{_{FT}}$ is described by the equation
\be
\ba{c}
\dps  z \bar z  - z\left(\frac{ \left(r_0 \left(r^2+1\right) \sin w_0 - \left(r^2_0+1\right) r \sin \phi \right)}{2 r_0 r \sin(\phi - w_0)} + i \frac{ \left(r_0 \left(r^2+1\right) \cos w_0 -\left(r^2_0+1\right) r \cos \phi \right)}{2 r_0 r \cos(\phi - w_0) }\right) -
\\
\\
\dps - \bar{z}\left(\frac{ \left(r_0 \left(r^2+1\right) \sin \psi-\left(r^2_0+1\right) r \sin \phi \right)}{2 r_0 r \sin(\phi -w_0)}- i \frac{ \left(r_0 \left(r^2+1\right) \cos \psi -\left(r^2_0+1\right) r \cos \phi \right)}{2 r_0 r \cos(\phi -w_0) }\right) + 1=0\;,
\ea
\ee
while the edges connecting points $z_{1,2}$ and  $z_{_{FT}}$ are described by the equations
\be
\ba{l}
\dps z \bar z  - \left(\frac{\left(r^2+1\right) \sin w_{1,2}-2 r \sin \phi}{2 r \sin (w_{1,2}-\phi ) } + i \frac{\left(r^2+1\right) \cos w_{1,2}-2 r \cos \phi}{2 r \sin (w_{1,2}-\phi ) }\right) z -
\\
\\
\dps
\hspace{30mm}- \left(\frac{\left(r^2+1\right) \sin w_{1,2} -2 r \sin \phi}{2 r \sin (w_{1,2}-\phi ) } - i \frac{\left(r^2+1\right) \cos w_{1,2}-2 r \cos \phi}{2 r \sin (w_{1,2}-\phi ) }\right) \bar z + 1 = 0\;.
\ea
\ee
The corresponding slope coefficients \eqref{slope} can be directly read off from these formulas,
\be
\ba{c}
\dps
\varkappa_1 =\frac{\sin w_1-r(r \sin (w_1 - 2 \phi )+2 \sin \phi )}{r (r \cos (w_1-2 \phi )-2 \cos \phi )+\cos w_1}\;,
\qquad
\varkappa_2 = \frac{\sin w_2-r(r \sin (w_2 - 2 \phi )+2 \sin \phi )}{r (r \cos (w_2-2 \phi )-2 \cos \phi )+\cos w_2}\;,
\\
\\
\dps
\varkappa_3 = \frac{\left(r^2_0+1\right) r \sin\phi-r_0 \left(r^2 \sin (2 \phi - w_0)+\sin w_0 \right)}{\left(r^2_0+1\right) r \cos \phi - r_0 \left(r^2 \cos (2 \phi - w_0 )+\cos w_0\right)}\;.
\ea
\ee
Now, we substitute the slopes into the equation system \eqref{slope_eq}. To this end,  using the \eqref{tangent} we write down cosines of the angles $\gamma_{21}, \gamma_{10}, \gamma_{20}$ as follows
\be
\label{cequ1}
\ba{c}
\dps \frac{\left(r^4+1\right) \cos (w_1-w_2)+2 r\left(r \cos (w_2 - w_1 -2 \phi )-\left(r^2+1\right) (\cos (w_1-\phi )+\cos (w_2-\phi ))\right)+4 r^2}{\left(r^2-2 r \cos (w_1-\phi )+1\right) \left(r^2-2 r \cos (w_2-\phi )+1\right)} = \cos\gamma_{21}
\ea
\ee
\be
\label{cequ2}
\ba{c}
\dps \frac{2 r \left(r \left(r^2_0 + r_0 \cos (w_1+ w_0 -2 \phi )+1\right)-r_0 \left(r^2+1\right) \cos (\phi - w_0 )\right) \cos (w_1-\phi )}{\left(r^2-2 r \cos (w_1-\phi )+1\right) \sqrt{\left(r^2_0-2 r_0 r \cos (\phi - w_0 )+r^2\right) \left(r^2_0 r^2-2 r_0 r \cos (\phi - w_0 )+1\right)}} -
\\
\\
\dps - \frac{-\left(r^2_0+1\right) r \left(r^2+1\right)-r_0 \left(r^4+1\right) \cos (w_1 - w_0 )}{\left(r^2-2 r \cos (w_1-\phi )+1\right) \sqrt{\left(r^2_0-2 r_0 r \cos (\phi - w_0 )+r^2\right) \left(r^2_0 r^2-2 r_0 r \cos (\phi -w_0 )+1\right)}} = \cos\gamma_{10}
\ea
\ee
\be
\label{cequ3}
\ba{c}
\dps \frac{2 r \left(r \left(r^2_0+ r_0 \cos (w_2+ w_0 -2 \phi )+1\right)-r_0 \left(r^2+1\right) \cos (\phi - w_0 )\right) \cos (w_2-\phi )}{\left(r^2-2 r \cos (w_2-\phi )+1\right) \sqrt{\left(r^2_0-2 r_0 r \cos (\phi - w_0 )+r^2\right) \left(r^2_0 r^2-2 r_0 r \cos (\phi -w_0 )+1\right)}} -
\\
\\
\dps - \frac{-\left(r^2_0+1\right) r \left(r^2+1\right)-r_0 \left(r^4+1\right) \cos (w_2- w_0 )}{\left(r^2-2 r \cos (w_2-\phi )+1\right) \sqrt{\left(r^2_0-2 r_0 r \cos (\phi - w_0 )+r^2\right) \left(r^2_0 r^2-2 r_0 r \cos (\phi - w_0 )+1\right)}}  = \cos\gamma_{20}
\ea
\ee
where the right-hand sides are given by  \eqref{cooos}. The resulting equation system is very complicated but it can be drastically simplified using the  regularized exponentiated lengths \eqref{regL}. Let us  introduce the following variables
\be
\ba{c}
\dps L_{1} =\frac{1+r^2 - 2r \cos(w_1-\phi)}{1-r^2}\;,
\quad
L_{2} = \frac{1+r^2 - 2r \cos(w_2-\phi)}{1-r^2}\;,
\quad
L_{0} = \frac{1+U}{1-U}\;,
\ea
\ee
where
\be
U=\sqrt{\frac{r^2_0-2 r_0 r \cos (\phi -w_0 )+r^2}{r_0^2 r^2-2 r_0 r \cos (\phi -w_0 )+1}}\;.
\ee
The first and second expressions are exponentiated lengths of the edges connecting $z_{1,2}$ and $z_{_{FT}}$, the third one is the exponentiated length of the edge connecting $z_{_{FT}}$ and $z_0$. Then, equations \eqref{cequ1}--\eqref{cequ3} take the form
\be
\label{LLL}
\ba{c}
\dps 1- \frac{2\sin^2  w_{21}}{L_{1}L_{2}} =  \cos \gamma_{12}\;,
\quad
\dps  \frac{K_2 -L_{2}(1+L^2_{0})}{L_{2}(L^2_{0}-1)} = \cos \gamma_{20}\;,
\quad
\dps  \frac{K_1 -L_{1}(1+L^2_{0})}{L_{1}(L^2_{0}-1)} = \cos \gamma_{10}\;,
\ea
\ee
where $ K_{1,2}$  are functions of initial coordinates,
\be
K_{1,2} = \frac{1+r^2_0 - 2 r_0 \cos(w_{1,2}-w_0)}{1-r^2_0}\;.
\ee
Remarkably, equations \eqref{LLL} are linear in $L_{1,2}$ and quadratic with respect to $(L_0)^2$, and, therefore, can be explicitly solved. Recalling the notation \eqref{new} we represent the solution as \eqref{final_solution_t}. Thus, the final answer is \eqref{masterN3}.

\paragraph{N=4 tree in Section \bref{sec:N4}.}  Let $\epsilon_{1}=\epsilon_{2}$, $\epsilon_{3}=\epsilon_{4}$ and  $\tilde \epsilon \neq \epsilon_{1,3}$. We claim that  \eqref{junction} is minimized with  respect to the junction point $z_0 = (r_0,w_0)$. Making use of  \eqref{masterN3} we can find the length function
\be
\label{exbr}
\ba{l}
\dps  L^{(4)}_{\disk}(w_1,w_2,w_3,w_4) = 2\epsilon_{1}\log\sin w_{21} + 2\epsilon_{3}\log\sin w_{43}\, -
\\
\\
\hspace{35mm}- \tilde{\epsilon}\log\left(\sqrt{P_{12}} - \sqrt{P_{12} - 1}\right)\left(\sqrt{P_{34}} - \sqrt{P_{34} - 1}\right)+  \tilde{C}\;,
\ea
\ee
where
\be
\label{B13}
\ba{c}
\dps
P_{ij} = \frac{K_i K_j}{\sin^2  w_{ij}}\;,
\qquad
K_i = \frac{1+r^2_0 -2 r_0\cos[w_i - w_0]}{1-r^2_0}\;,
\ea
\ee
$$
\ba{c}
\dps
\tilde{C} =  \frac{\tilde{\epsilon}}{2}\left(\log \frac{(\gamma_{1}-1)(\gamma_{3}-1)}{(\gamma_{1}+1)(\gamma_{3}+1)} + \gamma_{1} \log \frac{\gamma_{1}^2}{\gamma_{1}^2-1} + \gamma_{3} \log \frac{\gamma_{3}^2}{\gamma_{3}^2-1}\right),\;\; \gamma_{1} = \frac{2\epsilon_1}{\tilde{\epsilon}}\;,\;\gamma_{3} = \frac{2\epsilon_3}{\tilde{\epsilon}}.
\ea
$$

Let us extremize the function \eqref{exbr}. Evaluating first derivatives in $z_0$ we obtain the following two relations
\be
\label{minbr}
\noindent
\ba{l}
\dps \frac{\sin (\tilde{w}_{21}- w_0)}{\sqrt{r_0^2-2 r_0 \cos (w_1-w_0)+1}\sqrt{r_0^2-2 r_0 \cos (w_2-w_0)+1} } =
\\
\\
\hspace{30mm}\dps = \frac{\sin (-\tilde{w}_{43} + w_0)}{\sqrt{r_0^2-2 r_0 \cos (w_4 - w_0 )+1}\sqrt{r_0^2-2 r_0 \cos (w_3-w_0 )+1} }\;,
\\
\\
\ea
\ee
\be
\label{minbr2}
\ba{l}
\dps  \frac{2 r_0 \cos w_{21} - (1+r_0^2)\cos(\tilde{w}_{21} - w_0)}{\sqrt{r_0^2-2 r_0 \cos (w_1 - w_0 )+1}\sqrt{r_0^2-2 r_0 \cos (w_2- w_0 )+1} } =
\\
\\
\hspace{30mm} \dps = \frac{-2 r_0 \cos w_{43} + (1+r_0^2)\cos(\tilde{w}_{43} - w_0)}{\sqrt{r_0^2-2 r_0 \cos (w_4 - w_0)+1}\sqrt{r_0^2-2 r_0 \cos (w_3-w_0 )+1} }\;,
 \\
 \\

\ea
\ee
where we denoted $\tilde{w}_{ij} = (w_i + w_j)/2$. Solving this system of two equations one can fix  coordinates of the junction point $z_0 = (r_0,w_0)$. After that, substituting $z_0$ into the initial function \eqref{exbr} we will find a sought minimal total length. However, the equations are hard to integrate. Moreover, it turns out that there is a continuous family of roots.

Let us discuss a similar problem on  the Euclidean plane $\mathbb{R}^2$. Suppose that we want to find a point $z_0\in \mathbb{R}^2$ that minimizes the sum of  distances from $z_0$ to the points $(0, 0)$ and $(1,0)$ on the $x$-axis. The total length function is given by  $f(x_0,y_0) =  \sqrt{x_0^2 + y_0^2} + \sqrt{(x_0-1)^2 + (y_0 - 1)^2}$ and the minimization condition is  $d f =0$. One can explicitly show that a general solution is given by $z_0 = (x_0,0)$ for $\forall x_0 \in [0,1]$. Therefore, in order to  fix $x_0$ one is free to impose an additional condition consistent with the minimization conditions.

The analysis on $\disk$ is essentially the same and the junction point cannot be fixed unambiguously by two minimization equations \eqref{minbr} and \eqref{minbr2}. We choose an additional condition as
\be
\label{condbr}
K_1 K_2 = K_3 K_4\;.
\ee
It is consistent with \eqref{minbr} and \eqref{minbr2}. Then, the solution to \eqref{minbr}--\eqref{condbr} is given by
\be
\label{sb}
\ba{c}
\dps r_0 = \frac{1}{2} \left[\cos w_{21} + \cos w_{43} - \sqrt{(\cos w_{21} + \cos w_{43})^2 - 4\cos^2\frac{\dps \tilde{w}_{21} - \tilde{w}_{43}}{2}}\right]\sec \frac{\tilde{w}_{21} - \tilde{w}_{43}}{2}\;,
\ea
\ee
$$
w_0 = \frac{\tilde{w}_{43} + \tilde{w}_{21}}{2}\;.
$$
Substituting  \eqref{sb} into  \eqref{exbr} we obtain the final length function \eqref{answ_br}, \eqref{defU}.

\section{Proving the factorization relation}
\label{app:factor}

Here, we prove the factorization relation \eqref{factor} for any $s$. To simplify our presentation we introduce the notation $x_i = c_i(1-z_i)-\epsilon_i$. Then, the relation \eqref{ccc} is rewritten as
\be
\label{ccc_x}
\sum^{n-2}_{i=1}x_i =0\;.
\ee
The fusion rules around the identity exchange channel are given by \eqref{fusion_gen} (see Fig. \bref{fig:factor}). Our strategy below is to write down the monodromy system for the maximal $(s+3)$-point block and then use it in the monodromy system for the $s$-th identity $n$-point block. We will see that the $n$-point system decouples into two subsystems according to the factorization relation.

\paragraph{Maximal identity block.} Let us consider the maximal identity block $f^{(s)}_{s+3}(z^\prime|\epsilon^\prime, \tepsilon^\prime)$,  where $z^\prime =\{0, z_2,..., z_{s+1}\}$. By definition of the maximal identity block we have to consider first the monodromy equations for the general $(s+3)$-point block and then to impose the fusion condition \eqref{fusion_gen}. The $(s+3)$-point  monodromy equations \eqref{moneq}-\eqref{Is} take the form
\be
\label{vac}
I_{+-}^{(s+3|k)}\,I_{-+}^{(s+3|k)}+\left(I_{++}^{(s+3|k)}\right)^2  + 4\pi^2 \tilde\epsilon^2_{k}=0\;,\qquad k = 1,...\,, s\,,
\ee
where
\be
\label{IsApp}
\ba{c}
\dps
I^{(s+3|k)}_{+-}  =\frac{2\pi i}{\alpha}\left[\alpha \epsilon_1+\sum_{i=2}^{s+1}x_i-\sum_{i=2}^{k+1}(1-z_i)^\alpha(x_i-\alpha \epsilon_i)\right]\;,
\vspace{-3mm}
\\
\\
\dps
I^{(s+3|k)}_{-+}  = I^{(s+3|k)}_{+-}\big|_{\alpha \rightarrow -\alpha}\,,
\qquad\;\;
I^{(s+3|k)}_{++} = \frac{2\pi i}{\alpha}\sum_{i=k+2}^{s+1}x_i\,.
\ea
\ee
The accessory parameters $c_1, c_2,..., c_{s+1}$ satisfy the relation of the type \eqref{ccc} (see also \eqref{vac}),
\be
\label{ccc_vac}
\sum^{s+1}_{i=1}x_i =0\;.
\ee
Note that since $\tepsilon_s = 0$ and $I_{++}^{(s+3|s)} =0$, then  the $k=s$ equation in \eqref{vac}  factorizes as $I_{+-}^{(s+3|s)}\,I_{-+}^{(s+3|s)}=0$. The two factors read
\be
\label{linear_vac}
\ba{c}
\dps\alpha \epsilon_1+\sum_{i=2}^{s+1}x_i-\sum_{i=2}^{s+1}(1-z_i)^\alpha(x_i-\alpha \epsilon_i)=0\;,
\vspace{-3mm}

\\
\\
\dps-\alpha \epsilon_1+\sum_{i=2}^{s+1}x_i-\sum_{i=2}^{s+1}(1-z_i)^{-\alpha}(x_i+\alpha \epsilon_i)=0\;.
\ea
\ee
These relations can be substituted into $I^{(s+3|s-1)}_{+-}$ and $I^{(s+3|s-1)}_{-+}$. In particular, it follows that the $k=s-1$  equation in \eqref{vac} is satisfied identically because all terms are collected into two $(\pm)$ identical groups of differences of the squares. One concludes that two of the quadratic equations in \eqref{vac} are now replaced by two linear relations \eqref{linear_vac}.

\paragraph{Non-identity block.} Let us consider the second factor in \eqref{factor} which is $(n-s-1)$-point block $f^{(0)}_{n-s-1}(z^{\prime\prime}|\epsilon^{\prime\prime}, \tepsilon^{\prime\prime})$, where $z^{\prime\prime} = \{z_{s+2}, ..., z_{n-2}\}$. Note that here the first point $z_{s+2}\neq 0$ and, therefore, the respective monodromy equations take the properly relabeled  form \eqref{moneq}, \eqref{Is}, namely
\be
\label{gen_block}
I_{+-}^{(n-s-1|k)}\,I_{-+}^{(n-s-1|k)}+\left(I_{++}^{(n-s-1|k)}\right)^2  + 4\pi^2 \tilde\epsilon^2_{k}=0\;,\qquad k = s+2,...\,, n-3\,,
\ee
where
\be
\label{IsApp_gen}
\ba{c}
\dps
I^{(n-s-1|k)}_{+-}  =\frac{2\pi i}{\alpha}\left[\left(1-z_{s+2}\right){}^{\alpha }\left(\alpha \epsilon_{s+2}+\sum_{i=s+3}^{n-s-3}x_i\right)-\sum_{i=s+3}^{k+1}(1-z_i)^\alpha(x_i - \alpha \epsilon_i)\right],
\vspace{-3mm}
\\
\\
\dps
I^{(n-s-1|k)}_{-+}  = I^{(n-s-1|k)}_{+-}\big|_{\alpha \rightarrow -\alpha}\,,
\qquad\;\;
I^{(n-s-1|k)}_{++} = \frac{2\pi i}{\alpha}\sum_{i=k+2}^{n-2}x_i\,.
\ea
\ee

\paragraph{$s$-th identity block.} Now we consider the identity block $f^{(s)}_n(z|\epsilon, \tepsilon)$ on the left-hand side of the factorization relation \eqref{factor}. It is described by the $n$-point monodromy system \eqref{moneq}, \eqref{Is} with $z_1=1$ and the fusion constraints \eqref{fusion_gen} imposed.

It is obvious that the first $s$ equations of the system are given by the maximal identity block equations \eqref{vac} with accessory parameters $c_{2},..., c_{s+1}$. Then, from \eqref{ccc_x} and \eqref{ccc_vac} it follows that the remaining accessory parameters $c_{s+2},..., c_{n-2}$ satisfy the relation
\be
\label{ccc_p}
\sum^{n-2}_{i=s+2}x_i =0\;.
\ee

Let us consider now the $k=s+1$ equation of the $n$-point monodromy system \eqref{moneq}
\be
I_{+-}^{(n|s+1)}\,I_{-+}^{(n|s+1)}+\left(I_{++}^{(n|s+1)}\right)^2  + 4\pi^2 \epsilon^2_{s+2}=0\;,
\ee
where we used the fusion rule \eqref{fusion_gen}. Substituting \eqref{linear_vac} into $I_{\pm \mp}^{(n|s+1)}$  we can show that this equation is identically satisfied.

The other equations with $k=s+2,...$ of the system \eqref{moneq} are given by
\be
\label{others}
I_{+-}^{(n|k)}\,I_{-+}^{(n|k)}+\left(I_{++}^{(n|k)}\right)^2  + 4\pi^2 \tepsilon^2_{k}=0\;,
\qquad k=s+2, ..., n-3\;.
\ee
These equations are non-trivial and identical to \eqref{gen_block} and \eqref{IsApp_gen}. Indeed, taking account of \eqref{ccc_p} and then  \eqref{linear_vac} we obtain
\be
\ba{c}
\dps
I_{+-}^{(n|k)} = \frac{2\pi i}{\alpha}\left[\alpha \epsilon_1+\sum_{i=2}^{n-2}x_i-\sum_{i=2}^{k+1}(1-z_i)^\alpha(x_i-\alpha\epsilon_i)\right]
\\
\\
\dps\hspace{10mm}\approx  \frac{2\pi i}{\alpha}\left[\alpha \epsilon_1+\sum_{i=2}^{s+1}x_i-\sum_{i=2}^{k+1}(1-z_i)^\alpha(x_i-\alpha\epsilon_i)\right]
\vspace{-3mm}
\ea
\ee
$$
\approx \frac{2\pi i}{\alpha}\left[-(1-z_{s+2})^\alpha(x_{s+2} - \alpha \epsilon_{s+2})-\sum_{i=s+3}^{k+1}(1-z_i)^\alpha(x_i-\alpha\epsilon_i)\right]
$$
$$
\approx \frac{2\pi i}{\alpha}\left[(1-z_{s+2})^\alpha(\alpha \epsilon_{s+2}+\sum_{i = s+2}^{n-2}x_i)-\sum_{i=s+3}^{k+1}(1-z_i)^\alpha(x_i-\alpha\epsilon_i)\right]\;.
$$
The last expression is exactly $I_{+-}^{(n-s-1|k)}$ from \eqref{IsApp_gen}. On the other hand, $ I_{++}^{(n|k)}$ is of the required form as well. Therefore, we conclude that the equations \eqref{others} do describe an $(n-s-1)$-point block and the factorization condition \eqref{factor} is satisfied.

\providecommand{\href}[2]{#2}\begingroup\raggedright\endgroup

\end{document}